\documentclass[preprint,3p,authoryear,12pt]{elsarticle}
\usepackage{hyperref}
\usepackage{amssymb}
\usepackage{eqnarray,amsmath}
\usepackage{color}
\newcommand{\ket}[1]{\left| #1\right\rangle}

\journal{AAMOP}
\begin{document}
\begin{frontmatter}

\title{Optical Nanofibers: a new platform for quantum optics}

\author[label1]{Pablo Solano}
\author[label1]{Jeffrey A. Grover}
\author[label1]{Jonathan E. Hoffman}
\author[label1,label2]{Sylvain Ravets}
\author[label3]{Fredrik K. Fatemi}
\author[label1]{Luis A. Orozco}
\author[label1]{Steven L. Rolston}
\address[label1]{Joint Quantum Institute, Department of Physics, University of Maryland and NIST, College Park, MD 20742, USA.}
\address[label2]{Laboratoire Charles Fabry, Institut d'Optique, CNRS, Univ Paris Sud, 2 Avenue Augustin Fresnel, 91127 Palaiseau cedex, France}
\address[label3]{Army Research Laboratory, Adelphi, MD 20783, USA.}

\begin{abstract}
The development of optical nanofibers (ONF) and the study and control of their optical properties when coupling atoms to their electromagnetic modes has opened new possibilities for their use in quantum optics and quantum information science. These ONFs offer tight optical mode confinement (less than the wavelength of light) and diffraction-free propagation. The small cross section of the transverse field allows probing of linear and non-linear spectroscopic features of atoms with exquisitely low power. The cooperativity -- the figure of merit in many quantum optics and quantum information systems -- tends to be large even for a single atom in the mode of an ONF, as it is proportional to the ratio of the atomic cross section to the electromagnetic mode cross section. ONFs offer a natural bus for information and for inter-atomic coupling through the tightly-confined modes, which opens the possibility of one-dimensional many-body physics and interesting quantum interconnection applications.
The presence of the ONF modifies the vacuum field, affecting the spontaneous emission rates of atoms in its vicinity. The high gradients in the radial intensity naturally provide the potential for trapping atoms around the ONF, allowing the creation of one-dimensional arrays of atoms. The same radial gradient in the transverse direction of the field is responsible for the existence of a large longitudinal component that introduces the possibility of spin-orbit coupling of the light and the atom, enabling the exploration of chiral quantum optics. 
\end{abstract}


\end{frontmatter}

\tableofcontents
\addtocontents{toc}{\setlength{\parskip}{0pt}}
\newpage

\bibliographystyle{elsarticle-harv}

   \pagenumbering{arabic}
 
\section{Introduction}
\label{sec:introduction}

Quantum optics has followed a path to achieve the ideal limit of one single quantum of light, a photon, interacting with one single quantum of matter, an atom. The interest in this realization has theoretical and experimental implications that have illuminated and guided much of the contemporary discussion on quantum information.

In the case of photon-mediated interactions, the advent of cavity quantum electrodynamics (QED)~\cite{Haroche2006}, whereby cavities formed by mirrors or other structures modify the vacuum modes of the electromagnetic field while providing a preferential mode for the atom-light coupling, marked a transformative milestone. Cavity QED ushered in the ability to sufficiently isolate a quantum system from its environment and control nearly all of its degrees of freedom~\cite{Kimble1998}. This has led, for instance, to the demonstration of the Purcell effect -- increased or inhibited spontaneous emission rates~\cite{Hinds1994} -- and to generation of highly nonclassical photon states~\cite{Haroche2013}, among other phenomena. Remarkably, these studies have been realized within many different regions of the electromagnetic spectrum, e.g.\ from the microwave~\cite{Wallraff2004, Haroche2013} to optical domains~\cite{Kimble1994,Reiserer2015}. The emerging field of waveguide QED, which applies the machinery of cavity QED to propagating modes in electromagnetic structures, is growing and the topic of this review falls into this subject \cite{Calajo2016,Fang2015,Paulisch2016,Lodahl2015}.

Our study of atom-light interactions is framed within this context. We motivate the use of evanescent waveguides helped by the notion of cooperativity, which we define in the next paragraph.

%-------------------------------------------------------------------------------------------------------------------

\subsection{Cooperativity and optical depth}
\label{sec:coop}

Consider a two-level atom with dipole moment $\vec{d}$ interacting with an electric field $\vec{E}$ whose average energy is equivalent to one photon $({\hbar}\omega)$. The parameter, $g$, gives the strength of the coupling in frequency units, numerically equivalent to half of the vacuum Rabi splitting \cite{Sanchez1983},
\begin{equation}
\label{eq:dipole}
g = \frac{\vec{d} \cdot \vec{E}}{\hbar}\,.
\end{equation}
For an atom with decay rate, $\gamma$, and a field with decay rate, $\kappa$, we define the single-atom cooperativity to be~\cite{Kimble1998}
\begin{equation}
\label{eq:coop}
C_1 =  \frac{g^2}{\kappa\gamma}\,.
\end{equation}
A cooperativity of $C_1>1$ 
means that the rate that governs the interaction between the atom and the field mode is larger than the atomic and field reservoir coupling rates.
This places the system in the so-called strong coupling regime, which was a longstanding goal within the quantum optics community and that has been achieved in several systems, such as ions~\cite{Wineland2013}, Rydberg atoms in microwave cavities~\cite{Haroche2013}, neutral atoms in optical cavities~\cite{Kimble1994}, excitons in semiconductor microcavities~\cite{Gibbs2011}, optomechanical systems \cite{Groblacher2009} and superconducting circuits in planar waveguides~\cite{Blais2004}.

To better understand how one can coerce a system into the strong coupling regime where a single mode of the electromagnetic field and one atom preferentially exchange an excitation, it can be useful to relate the cooperativity to the optical depth $(OD)$. This way of thinking comes from the predecessor of cavity QED, optical bistability~\cite{Lugiato1984}, which is part of the more general area of dissipative systems in quantum optics as treated by~\cite{Bonifacio1982}. We illustrate the argument by considering a high-finesse Fabry-Perot cavity with mirror transmission $T$ and mirror separation length $L$ so that its full width at half maximum (FWHM) is $2\kappa$, with $\kappa = cT/2L$ as a frequency half-width. The electric field amplitude for a field with an average energy of a single photon within this mode is given by
\begin{equation}
\label{eq:efield}
E = \sqrt{ \frac{\hbar \omega}{2\varepsilon_0 V} }\,,
\end{equation}
defining the mode volume to be $V = A_{\mathrm{mode}}\times L$. The free-space decay rate of the atom from Fermi's golden rule is
\begin{equation}
\label{eq:FGR}
\gamma_{0}= \frac{4 \omega^3}{3 c^2} \frac{d^2}{4 \pi \varepsilon_0 \hbar c}\,,
\end{equation}
where $d$ is the magnitude of the dipole moment of the atom and $\omega = 2\pi c/\lambda$ is the resonant angular frequency of the decay transition associated with the wavelength $\lambda$ . Typical values of $d$ for the D2 line of alkali atoms are about 5$a_0e$, where $a_0$ is the Bohr radius and $e$ the electron charge~\cite{Safronova2004}. Combining Eq.~\ref{eq:efield} and Eq.~\ref{eq:FGR} into Eq.~\ref{eq:coop} yields a single-atom cooperativity of~\cite{Tanji-Suzuki2011}
\begin{equation}
\label{eq:coop2}
C_1 = \frac{A_{\mathrm{atom}}}{A_{\mathrm{mode}}} \frac{1}{T}\,.
\end{equation}
Here we have defined the ``area'' of the atom, $A_{\mathrm{atom}}$, to be the resonant scattering cross section $\sigma_0 = 3 \lambda^2/2\pi$. This motivates a geometric framework to think about the cooperativity by realizing that the $OD$ for a dipole transition is:
\begin{equation}
\label{eq:OD}
OD=\frac{A_{\mathrm{atom}}}{A_{\mathrm{mode}}}\,.
\end{equation}
We conclude from Eq.~\ref{eq:coop2} that $C_1$ is just the product of the optical density of a single atom times the cavity enhancement factor $1/T$. 

Equation~\ref{eq:coop2} states that the cooperativity is independent of the atomic dipole moment and the cavity length $L$, but depends on the overlap of the two ``areas''. See~\cite{Bonifacio1982} for a discussion not only of resonance fluorescence, but also optical bistability and superradiance and their connection through the idea of cooperativity. Efforts to increase this figure of merit have followed a few different paths. 

For certain processes with $N$ atoms, the total cooperativity, $C=C_{1}N$, is important. Increasing $N$ achieves an appropriate threshold of the system, {\it e. g.} vapor cells with high atomic densities facilitate the observation of coherent processes such as electromagnetically induced transparency (EIT)~\cite{Fleischhauer2005}.

The Nobel Prize-worthy efforts of Serge Haroche focused on decreasing $T$ with microwave cavities possessing finesses greater than $10^9$ while making sure that the cavity-mode cross section significantly overlaps with the properly aligned Rydberg atom cross section. With this system, his group created highly nonclassical states and performed quantum non-demolition measurements of photon jumps~\cite{Haroche2013}.

Recent advances in superconducting technology have allowed physicists to create nonlinear quantum circuits that behave like ``artificial atoms''~\cite{Makhlin2001}. By coupling these so-called qubits to a high-quality-factor coplanar resonator, scientists have engineered an analog of cavity QED, dubbed circuit QED, that achieves couplings far beyond what have been realized in optical systems~\cite{Blais2004, Wallraff2004}. This architecture not only relies on the high finesse of the cavity to increase $C$, but the area of the artificial atoms (antennae, qubits) has also been increased significantly beyond that of the mode. This limit can not yet be realized with atoms in free space, but may be achievable for atoms near photonic and plasmonic structures, where the field can be confined beyond the diffraction limit.

Finally, we mention recent efforts made by some groups which have moved away from the use of a traditional cavity altogether, trying to increase the cooperativity of an atom in free space, making the enhancement factor $1/T=1$ as in Eq.~\ref{eq:coop2}. One possibility is to use high-NA optical systems to focus light to a small spot and achieve high coupling in free space~\cite{Tey2008, Hetet2011, Streed2012}. Another possibility is to use a parabolic mirror that focuses a laser such that the focused beam has the same structure as the dipole radiation pattern of a single atom, thereby increasing the ratio of the atomic area to the mode area~\cite{Stobinska2009, Golla2012, Heugel2012}.

\begin{figure}
\centering
\includegraphics[width=1\textwidth]{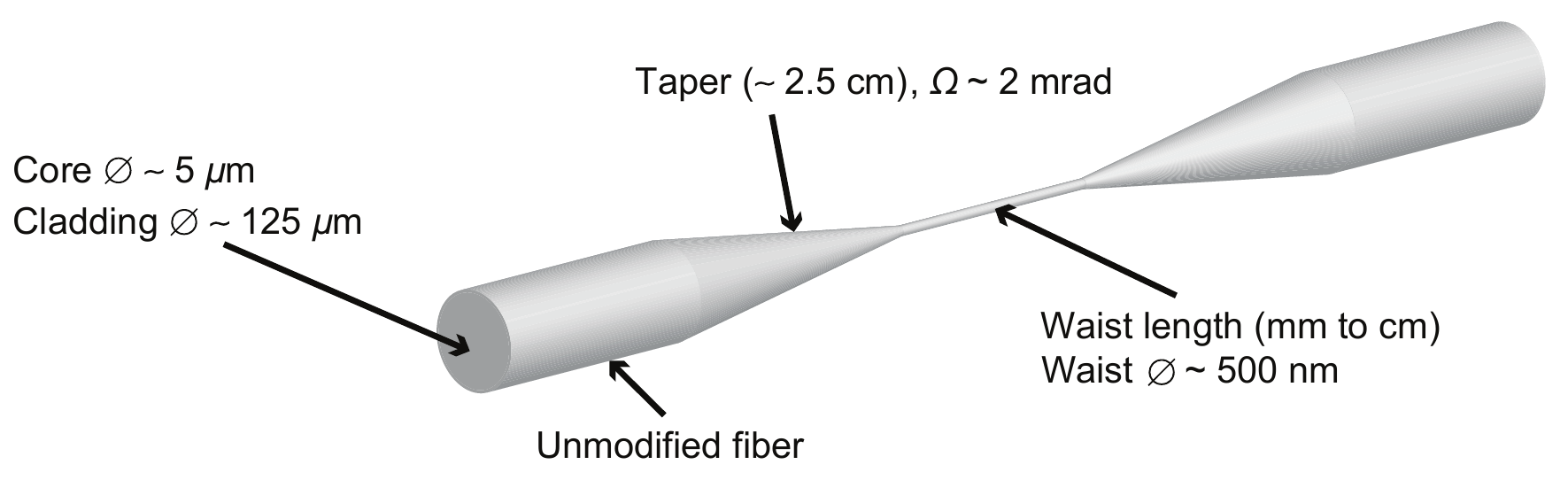}
\caption{Schematic of an optical nanofiber where the transverse dimension has been greatly expanded compared to the longitudinal. Atoms are around the waist region either free or trapped. There are three distinctive sections of the ONF and the typical values are those used in our experiments~\cite{Grover2015}.}
\label{fig:nanofiber}
\end{figure}

We will consider, in this review, the coupling of atoms to one kind of nanophotonic waveguide: an optical nanofiber (ONF). Nanophotonic waveguides are not like the traditional optical cavities with high finesses discussed above, but they do modify the vacuum mode structure in a nontrivial way. As a result of this modification and the appreciable overlap between the atomic and optical areas, one can couple an atom to the electric field of a photon in a single pass. In fact, there is an active area of research studying waveguides constructed via nanofabrication techniques, whose mode produces a large $OD$ for a single atom~\cite{Thompson2013, Goban2014, Goban2015, Hood2016}. Important advances have happened, for example, with hollow-core fibers: encasing an atomic vapor into the hollow core of a photonic-crystal fiber to confine atoms and
photons in the waveguide increases $C$, but the manipulation of the atoms is not as straightforward as if they are outside the photonic structure as in~\cite{Ghosh2006, Bajcsy2009, Venkataraman2011, Sprague2014}. Optical nanofibers (ONFs) formed by thinning single-mode optical fibers to sub-wavelength diameters, as shown in Fig.~\ref{fig:nanofiber} (not to scale), are another example of this kind of structure. It has been demonstrated that ONFs provide an excellent platform to interface trapped atoms to the evanescent field of the mode around a nanometer-size waist region~\cite{Vetsch2010, Goban2012, Beguin2014a, Lee2014,Kato2015,Corzo2016}. We review the platform of ONFs with atoms and their implications and applications for quantum physics.

%-------------------------------------------------------------------------------------------------------------------

\subsection{Nanofiber platform}

Before embarking on a thorough discussion of the nanofiber platform it is important to point out the advantages over other nanophotonic structures that we see. 

Nanofibers can be produced in-house, using a heat-and-pull method~\cite{Hoffman2014a,Ward2014}. The glass malleability ensures  low surface roughness. The smoothness of the surface is a great asset since it leads to ultra-high transmission structures that can withstand high optical powers (almost one Watt in vacuum~\cite{Hoffman2014a}) without permanent damage to the fiber or degradation of the transmission.

Optical fibers also show great versatility in terms of connectivity to other systems. The advanced state of fiber optic technology is an enormous advantage to pursuing quantum information devices on this platform as they facilitate the interaction and communication among different modes and modular devices as stated by~\cite{Kimble2008}.
\cite{Kurizki2015,Xiang2013} review the growing area of hybrid quantum systems that are increasing in importance in quantum optics and quantum information. They combine different kinds of systems -- {\it e. g.} atomic, Rydberg, ionic, photonic, condensed matter -- to utilize the  best coherence available for different tasks: processing or memory. \cite{Hoffman2011} and \cite{Kim2011} propose the use  of a trapped atoms around an ONF to couple them through their magnetic dipole them to a  superconducting  circuit in a cryogenic environment. \cite{Hafezi2012} studies for such a system the atomic interface between microwave and optical photons. 

One of the most fascinating developments is the use of ONFs in quantum optics is for the study of chiral quantum optics~\cite{Lodahl2016} and its connections with many-body physics. ONFs indeed provide a unique platform to study this nascent area.

One difficulty with nanofibers comes from 
the polarization structure of the modes and its limited control along the waist length. However they have not been a major drawback for experiments. Also, the values of atom coupling to the nanofiber currently do not reach those recently seen in nanophotonic devices~\cite{Yu2014,Goban2015,Hood2016}, but the entire parameter space for traps has yet to be explored, and improvements may be possible.

%-------------------------------------------------------------------------------------------------------------------

\subsection{Cooperativity in a nanofiber}

We now present, following closely the discussion in~\cite{Quan2009} related to waveguide QED with atoms, the connections between emission enhancement, coupling efficiency, cooperativity, and the Purcell effect. Two quantities characterize the coupling between the nanofiber mode and the atomic dipoles around the fiber but inside the mode.

The emission enhancement parameter $\alpha$:
\begin{equation}
\label{eq:alpha}
 \alpha=\frac{\gamma_{{1D}}}{\gamma_{0}} 
\end{equation}
is the ratio between the emission rate into the quasi one-dimensional (1D) mode $\gamma_{{1D}}$ of the nanofiber in both possible directions combined and the intrinsic spontaneous emission rate of an atom in free space $\gamma_{0}$ (see Eq.~\ref{eq:FGR}),

The waveguide coupling efficiency $\beta$:
\begin{equation}
\label{eq:beta}
\beta=\frac{\gamma_{{1D}}}{\gamma_{{Tot}}}
\end{equation}
is the ratio between the emission rate into the waveguide mode in both directions combined and total emission rate into all radiative channels $\gamma_{{Tot}}$. 

The parameter $\alpha$ is proportional to the interaction rate of the atom and the single mode, while $\beta$ quantifies the fraction of the total rate that couples to the single mode. As an example to illustrate the two concepts consider an atom in a cavity with  dimensions less than $\lambda/2$. The system  has  $\alpha=0$ and $\beta=1$,  implying that there will be no decay signal despite $\beta=1$.

The single-atom cooperativity in terms of these parameters is
\begin{equation}
\label{eq:Cbeta}
C_1=\frac{\beta}{(1-\beta)}=\frac{\gamma_{1D}}{\gamma_{Tot}-\gamma_{1D}}\,.
\end{equation}
This expression relates the parameter $\beta$, often used in the microcavity literature \cite{Chang1996}, to $C_1$ and thus to what we have presented above. 
The meaning of Eq.~\ref{eq:Cbeta} in terms of rates is equivalent to  to Eq.~\ref{sec:coop} in the sense that $C$ is the ratio of the rate of interaction between atom and field mode to the interaction to other reservoirs.
The Purcell factor, the total enhancement of spontaneous emission, is the ratio $\alpha/\beta=\gamma_{{Tot}}/\gamma_{0}$. It is noteworthy that, in the presence of a nanofiber with index of refraction $n$, $C_{1}$ (connected to the energy density) is directly related to the $OD$ (connected to the energy flux) via the index of refraction, as we learn from classical electricity and magnetism~\cite{Blow1990}. Furthermore $C$ is related to the  $OD$ through an enhancement facttor, $1/T$, in cavity QED and for the nanofibers $[\gamma_0/(\gamma_{Tot}-\gamma_{1D})]$.

%-------------------------------------------------------------------------------------------------------------------

\subsection{Optical nanofibers as enablers of high cooperativity and optical depth }
\label{subsec:nanofibercoop}

Fiber optics in present day telecommunications have been a game-changing technology allowing enormous bandwidth for current and future uses. It began with the pioneering observation of the low-loss properties of glass fibers in~\cite{Kao1966}. This was recognized with the 2009 Nobel Prize awarded to Charles K. Kao. (See~\cite{Kao2010}). 

More recently, ONFs have seen widespread use in science and engineering~\cite{Brambilla2010, Chen2013,Morrissey2013}. The tight confinement of light around ONFs~\cite{LeKien2004OC}, unique geometries provided by the fiber modes~\cite{Sague2008b, LeKien2004, Reitz2012}, low loss, and promise of improved atom-light interaction~\cite{Alton2011, LeKien2005, Vetsch2010, Goban2012, Wuttke2012} have led to increased interest in the physics community. Optical micro- or nanofibers are used for sensing and detection~\cite{Knight1997, Nayak2007}, and coupling light to resonators~\cite{Knight1997, Kakarantzas2002, Spillane2003, Louyer2005, Morrissey2009, Fujiwara2012}, NV centers~\cite{Schroder2012}, or photonic crystals~\cite{Thompson2013,Sadgrove2013}.

Reducing the thickness of an optical fiber to sub-wavelength diameters~\cite{Tong2003} modifies the  boundary conditions of the field so that a significant fraction of the light propagates in an evanescent field around the fiber waist. Nanofibers thus provide an excellent platform to interface light with atoms. To confine atoms along the nanofiber, one can couple laser beams into the fiber to create an optical dipole potential around the waist~\cite{LeKien2004}. Optical dipole trapping of atoms is a well-developed technique applied to numerous atomic species. See the excellent review of~\cite{Grimm2000} for details on the requirements depending on specific magnetic sub-states and polarizations used.

Typical trapping schemes allow trap depths of fractions of a milliKelvin located a few hundred nanometers from the fiber (\cite{Vetsch2010, Goban2012, Beguin2014a,Lee2014, Kato2015,Corzo2016}). Trapping lifetimes of tens of milliseconds and coherence times of $\sim600\,\mu$s~\cite{Reitz2013} have been obtained and formation of a one-dimensional lattice along the nanofiber waist has also been demonstrated. In this regime, the $OD$ per atom is as large as a few percent, so that a modest atom number can achieve a large optical depth. This confirms that ONFs are a viable platform for studying the physics of light-matter interactions, as they enable high optical depth and cooperativity.

%-------------------------------------------------------------------------------------------------------------------

\subsection{Outline of review}
\label{ssec:outline}

We start the review in Sec. \ref{sec:modes} with a treatment of the nanofiber electromagnetic modes and show intensity profiles with particular attention to the polarization properties of these modes. Sec. \ref{sec:fabrication} details the fabrication and characterization of the nanofibers. We then in Sec.~\ref{sec:around} concentrate on recent studies with untrapped atoms around nanofibers, including spectroscopy. Current implementations of atom trapping around a nanofiber and some interesting experiments with such 1D arrays are presented in Sec.~\ref{sec:trapping}. Experiments in quantum optics and quantum information come in Sec.~\ref{sec:qo} including realizations of EIT, optical memories, and cavity QED. Sec.~\ref{sec:chirality}, the last section, discusses how nanofibers are currently contributing to the emergence of the field of chiral quantum optics. The review ends with concluding remarks.

\newpage
  % Fredrik update Jan 7 2017.  Still in progress
\section{Nanofiber electromagnetic modes}

\label{sec:modes}

A standard step-index optical fiber is a cylindrical waveguide having a light-guiding core of typical radius of $2-3~\mu$m and refractive index $n_1$, surrounded by a cylindrical cladding with index $n_2$.  In commercial, single-mode fibers, the refractive index difference is small, with $0.001\lesssim n_1-n_2 \lesssim 0.02$, and the waveguide is considered to have ``weakly guiding'' core-cladding guidance.  As shown schematically in Fig.~\ref{fig:nanofiber}, as the fiber is tapered down this geometry adiabatically transforms to a step-index waveguide in which the light is entirely guided with air (or vacuum) as the cladding (``cladding-air'' guidance), having an index difference close to 0.5.  Detailed vector-mode solutions are required for such ``strongly guiding'' waveguides, and are described in a number of treatments~\cite{Marcuse1981,Snyder1983,Yariv1990,Marcuse1991, Sague2008a, Vetsch2010a}, while in the weakly guiding limit, a scalar treatment is sufficient.  Here we present a summary of the vector-mode solutions from~\cite{Hoffman2014}, and discuss some of the relevant aspects of optical nanofibers.

%=================================================================

\subsection{Field equations}
\label{sec:fieldeqs}

The electric and magnetic fields, \textbf{E} and \textbf{H}, are shown below in cylindrical coordinates for a single-step index geometry with core radius, $a$, of index $n_1$, and an infinite cladding of index $n_2$. The solution of the Maxwell Equations in cylindrical coordinates leads to the following expressions for the field components along the radial ($r$), azimuthal ($\phi$), and longitudinal ($z$, propagation) directions inside ($r<a$) and outside ($r>a$) the core.

For $r<a$:

\begin{align}
\label{eq:fieldsa}
E_{r, \pm} &=  \frac{-i \beta}{h^2} \left [\pm \frac{i \mu_0 \omega l}{\beta r} B J_l (hr) + A h J'_l (hr)  \right ] \mathrm{e}^{i (\omega t - \beta z \pm l \varphi)} \\
E_{\phi, \pm} &= \frac{-i \beta}{h^2} \left [ \pm \frac{i l}{r} A J_l (hr) - \frac{\mu_0 \omega h}{\beta} B J'_l (hr)  \right ] \mathrm{e}^{i (\omega t - \beta z \pm l \varphi)}\\
E_{z, \pm} &= A J_l (hr) \mathrm{e}^{i (\omega t - \beta z \pm l \varphi)}\\
H_{r, \pm} &= \frac{-i \beta}{h^2} \left [ \mp \frac{i \varepsilon_1 \omega l}{\beta r} A J_l (hr) + B h J'_l (hr)  \right ] \mathrm{e}^{i (\omega t - \beta z \pm l \varphi)}\\
H_{\phi, \pm} &= \frac{-i \beta}{h^2} \left [ \frac{\varepsilon_1 \omega}{\beta} A h J'_l (hr) \pm \frac{i l}{r} B J_l (hr)  \right ] \mathrm{e}^{i (\omega t - \beta z \pm l \varphi)}\\
H_{z, \pm} &= B J_l (hr) \mathrm{e}^{i (\omega t - \beta z \pm l \varphi)},
\end{align}

and $r>a$,

\begin{align}
\label{eq:fieldsr}
E_{r, \pm} &=\frac{i \beta}{q^2} \left [\pm \frac{i \mu_0 \omega l}{\beta r} D K_l (qr) + C h K'_l (qr)  \right ] \mathrm{e}^{i (\omega t - \beta z \pm l \varphi)}\\
\label{eq:fieldsphi}
E_{\phi, \pm} &= \frac{i \beta}{q^2} \left [ \pm \frac{i l}{r} C K_l (q r) - \frac{\mu_0 \omega h}{\beta} D K'_l (q r)  \right ] \mathrm{e}^{i (\omega t - \beta z \pm l \varphi)}\\
\label{eq:fieldsz}
E_{z, \pm} &= C K_l (qr) \mathrm{e}^{i (\omega t - \beta z \pm l \varphi)}\\
H_{r, \pm} &= \frac{i \beta}{q^2}  \left [ \mp \frac{i \varepsilon_2 \omega l}{\beta r} C K_l (qr) + D q K'_l (qr)  \right ]  \mathrm{e}^{i (\omega t - \beta z \pm l \varphi)}\\
H_{\phi, \pm} &= \frac{i \beta}{q^2} \left [ \frac{\varepsilon_2 \omega}{\beta} C q J'_l (qr) \pm \frac{i l}{r} D K_l (qr)  \right ]  \mathrm{e}^{i (\omega t - \beta z \pm l \varphi)}\\
H_{z, \pm} &= D K_l (qr) \mathrm{e}^{i (\omega t - \beta z \pm l \varphi)}\,,
\end{align}
where $\beta$ is the mode propagation constant, $h = \sqrt{k^2 - \beta^2}$, $q = \sqrt{\beta^2 - k^2}$, $k=2\pi/\lambda$ is the wavenumber, and $\varepsilon_i$ gives the dielectric constant in regions $i = {1,2}$.
The parameter $l$ is a non-negative integer that gives the order of the guided mode and its angular momentum.
We also use the notation $J'_l (hr) = \partial J_l (hr)/\partial (hr)$, $K'_l (qr) = \partial K_l (qr)/\partial (qr)$ for derivatives of Bessel functions $J_l$ and modified Bessel functions of the second kind $K_l$ of order $l$. 

Boundary conditions impose the following relations for the interior constants $A$ and $B$, and for the exterior constants $C$ and $D$:
\begin{align}
\label{eq:bcs}
\frac{B}{A} &= \pm \left [ \left( \frac{1}{ha} \right )^2 + \left ( \frac{1}{qa}  \right )^2  \right ] \,   \left [ \frac{J'_l (ha)}{ha J_l(ha)} + \frac{K'_l (qa)}{qa K_l (qa)}   \right ]^{-1} \\
\frac{C}{A} &= \frac{J_l (ha)}{K_l (qa)} \\
\frac{D}{A} &= \frac{B}{A} \frac{J_l (ha)}{K_l (qa)} - \frac{l^2 \beta^2}{k^2_0} \left [ \left( \frac{1}{ha} \right )^2 + \left ( \frac{1}{qa}  \right )^2  \right ]^2\,,
\end{align}
so that the knowledge of the propagation constant $\beta$ and amplitude normalization constant $A$ (Sec.~\ref{sec:norm}) completely define the system.

%=================================================================

\subsection{Propagation constant}
\label{sec:props}

\begin{figure}
\centering
\includegraphics[width=0.55\textwidth]{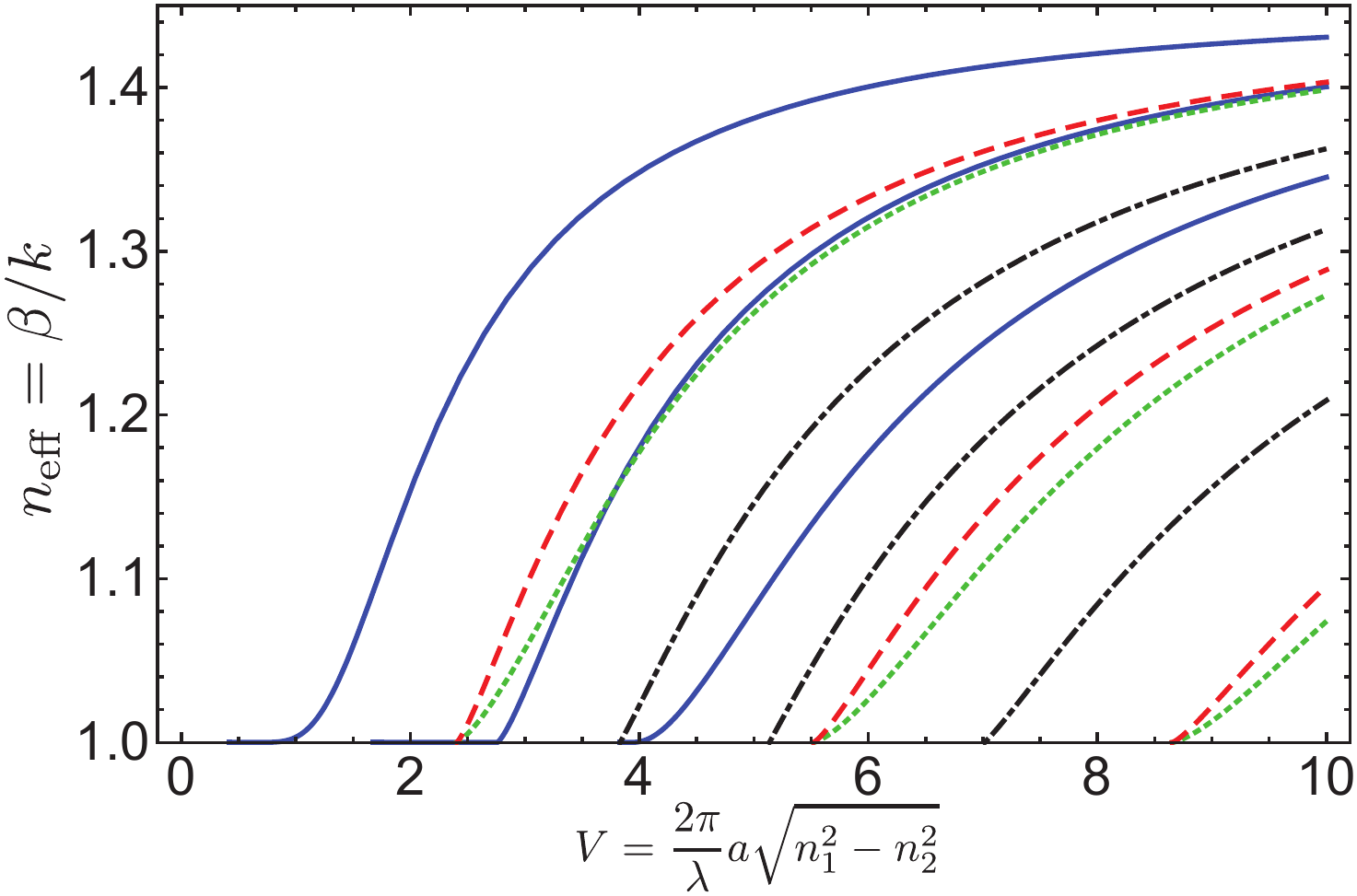}
\caption[Effective index of refraction as a function of V-number]{\label{fig:neff}Effective index of refraction as a function of V-number. The families of modes and their colors are $\mathit{HE}$ (solid blue), $\mathit{EH}$ (dashed-dotted black), $\mathit{TE}$ (dashed red), $\mathit{TM}$ (dotted green). (Reprinted from Fig. 4.1 with
permission from  Hoffman, J. E., 2014. Optical nanofiber fabrication and analysis towards coupling atoms to superconducting
qubits. Ph.D. thesis, University of Maryland College Park. Copyright (2014) J. E. Hoffman~\cite{Hoffman2014}).}
\end{figure}

An eigenvalue equation determines the propagation constant:
\begin{equation}
\label{eq:props}
 \frac{J_{l-1} (ha)}{ha J_l(ha)}  = \frac{\left (n^2_1 +n^2_2 \right )}{4 n^2_1} \left [  \frac{K_{l-1} (qa) + K_{l+1} (qa)}{qa K_l(qa)}  \right ] + \frac{l}{(ha)^2} \pm R,
\end{equation}
where
\begin{equation}
\label{eq:propsR}
 R = \sqrt{\frac{\left (n^2_1 - n^2_2 \right )^2}{\left ( 4 n^2_1 \right )^2} \left [  \frac{K_{l-1} (qa) + K_{l+1} (qa)}{qa K_l(qa)}  \right ]^2 + \frac{l^2 \beta^2}{n^2_1 k^2_0} \left [ \left( \frac{1}{ha} \right )^2 + \left ( \frac{1}{qa}  \right )^2  \right ]^2 }\,,
\end{equation}
and the $\pm R$ solutions correspond to $\mathit{EH}$ and $\mathit{HE}$ modes, respectively.
A normalized frequency called the $V$-number is defined by the relation 
\begin{equation}
\label{eq:V}
V =  \frac{2\pi}{\lambda} a \sqrt{n^2_1 - n^2_2}, 
\end{equation}
which scales the optical frequency by the fiber radius and its numerical aperture ($\sqrt{n^2_1 - n^2_2}$).
We can numerically solve Eq.~\ref{eq:props} for a particular $V$-number (Eq.~\ref{eq:V}) and $l$ by finding the points of intersection of its left hand side and right hand side.
Modes are labeled with subscripts $lm$, e.g. $\mathit{HE}_{lm}$, where for a given $l$, the successive points of intersection signify increasing $m$.
Fig.~\ref{fig:neff} plots the result of this calculation (where $n_{\mathrm{eff}} = \beta/k$) as a function of $V$ for various families of modes.
Note that the cutoff occurs at $V=2.405$ and that the fundamental $\mathit{HE}_{11}$ mode propagates for any $V>0$.  In the weak guiding limit, these vector mode solutions, $\mathit{HE}_{lm}$, become degenerate in the linearly-polarized (LP) basis, meaning that they have the same propagation constant.

%=================================================================

\subsection{Normalization}
\label{sec:norm}

The last parameter to determine is $A$, which is calculated using energy conservation.
We normalize the time-averaged Poynting vector in the z-direction relative to the input power,
\begin{equation}
\label{eq:poynting}
P = \langle S_z \rangle_t = A^2 \pi \left ( D_{\mathrm{in}} + D_{\mathrm{out}} \right )\,,
\end{equation}
where $D_{\mathrm{in}}$ and $D_{\mathrm{out}}$ will be found analytically.
For the $\mathit{HE}_{lm}$ and $\mathit{EH}_{lm}$ modes, these parameters are
\begin{align}
\label{eq:Ds}
D_{\mathrm{in}} &= \frac{\pi a \beta^2}{4 \mu_0 \omega} \frac{\beta}{h^2}  [ (1 + sl)(N^2_1 + sl)  [ J^2_{l+1} (ha) - J_l (ha) J_{l+2} (ha)  ] \nonumber \\
			& \qquad {} + (1-sl)(N^2_1 - sl)  [ J^2_{l-1} (ha) - J_l (ha) J_{l-2} (ha)  ] ] \\
D_{\mathrm{out}} &= \frac{-\pi a \beta^2}{4 \mu_0 \omega} \frac{\beta}{q^2} \left (\frac{J_l (ha)}{K_l (qa)} \right )^2 [ (1 + sl)(N^2_2 + sl)  [ K^2_{l+1} (qa) - K_l (qa) K_{l+2} (ha)  ] \nonumber \\
			& \qquad {} + (1-sl)(N^2_2 - sl)  [ K^2_{l-1} (qa) - K_l (qa) K_{l-2} (qa)  ] ] \,,
\end{align}
where $N_i = n_i k/\beta$ and $s = B \mu_0 \omega/(i l\beta)$ (with $B$ given by Eq.~\ref{eq:bcs}).

\begin{figure}
\centering
\includegraphics[width=0.75\textwidth]{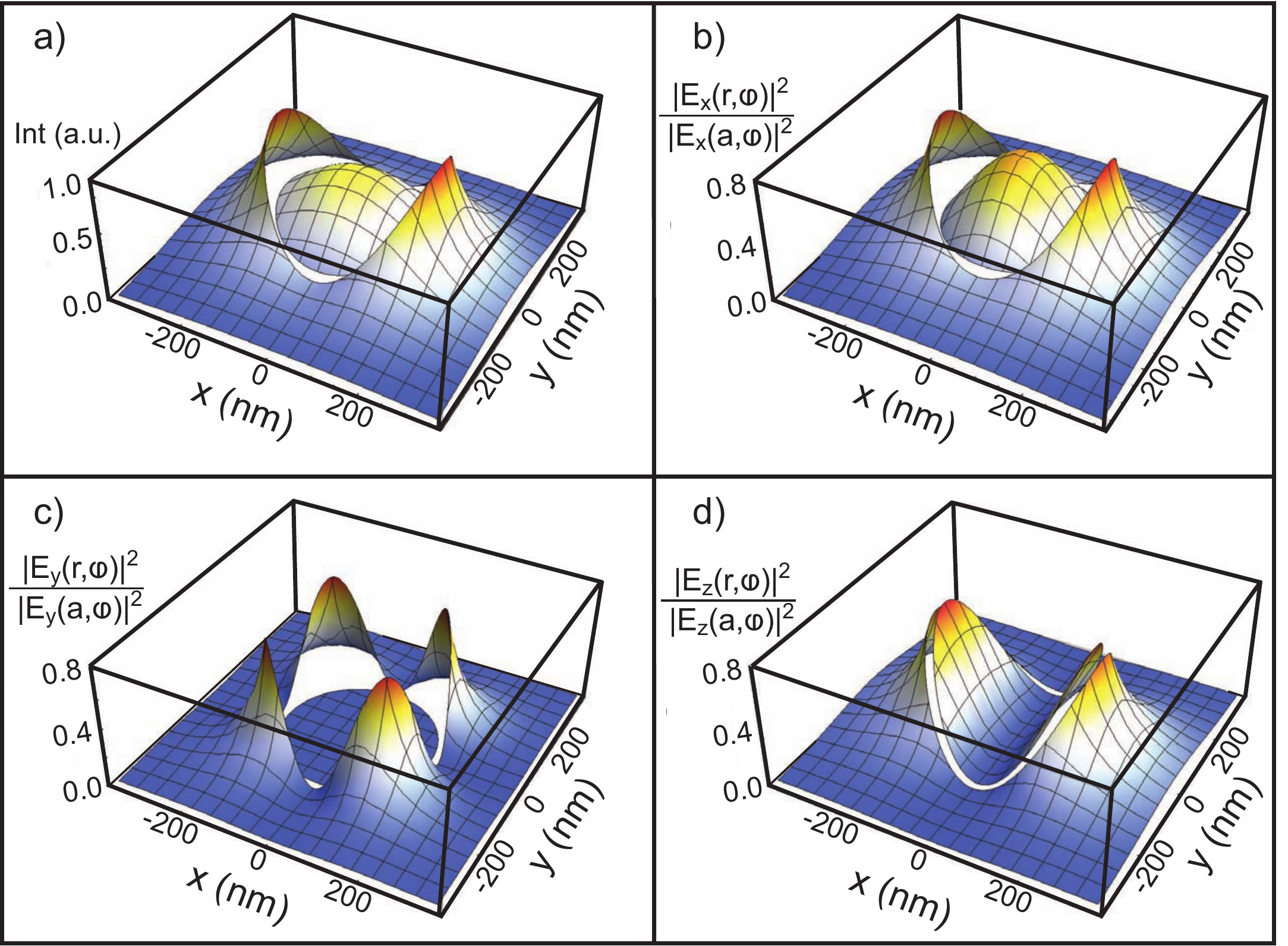}
\caption{\label{fig:HE11}Fundamental ($\mathit{HE}_{11}$) mode structure of 360-nm diameter nanofiber.  A Total intensity normalized to its maximum value. B Intensity of the $x$ component normalized to its maximum value at the nanofiber radius ($a$). C Intensity of the $y$ component normalized to its maximum value at the nanofiber radius ($a$).
D Intensity of the $z$ component normalized to its maximum value at the nanofiber radius ($a$).(Reprinted from Fig. A.7 with
permission from  Hoffman, J. E., 2014. Optical nanofiber fabrication and analysis towards coupling atoms to superconducting
qubits. Ph.D. thesis, University of Maryland College Park. Copyright (2014) J. E. Hoffman~\cite{Hoffman2014}).}
\end{figure}

Figure~\ref{fig:HE11} offers a summary  by plotting the mode structure of the $\mathit{HE}_{11}$ mode, showing the intensity (Fig. ~\ref{fig:HE11} a)) as well as the norm of each electric field component normalized to their value at the fiber surface (Fig. ~\ref{fig:HE11} b), c) and d)).
These values correspond to a 360-nm-diameter fiber with index of refraction $n_1 = 1.45367$ and propagating wavelength of 780~nm.
Note that the intensity has a sharp discontinuity at the fiber surface. The two largest components, $E_x$ and $E_z$, are comparable.  The presence of a sizable longitudinal component $E_z$ plays a critical role in nanophotonic atom-photon interactions~\cite{LeKien2004}. 

The radial decay of the evanescent field amplitude is not a simple exponential, but a complicated sum of modified Bessel functions of the second kind $K_{l}(qr)$ (see Eqs. \ref{eq:fieldsr}, \ref{eq:fieldsphi} and \ref{eq:fieldsz}). Since $r$ is the distance from the center of the ONF and we are interested in the field outside the dielectric media, the asymptotic expansion for large argument
$K_{l}(qr)\approx \sqrt{\pi/ 2 qr}e^{-qr}$
is a good approximation, for any order $l$. Considering this, the radial dependence of the evanescent electric field is
\begin{equation}
\label{eq:evanescent}
E_{i}(qr)\approx c_{i}r^{-\frac{1}{2}}e^{-qr},
\end{equation}
where $E_i$ is the $i$-th component of the electric field with amplitude proportional to $c_i$. This shows that the evanescent field decays at a shorter distance than an exponential decay. We have tested this approximation against exact numerical calculation showing excellent agreement, providing a simpler and more intuitive mathematical expression for the evanescent field. 

\subsection{Quasilinear polarization}

 When we launch linearly polarized light into a fiber it will excite both the $\pm l \phi$ solutions. 
 We can represent the mode in the quasilinear basis, as represented below.
\begin{align}
\bf{E}_{lin} &= \frac{1}{\sqrt{2}}\left[ \bf{E}_+ \pm \bf{E}_-\right] \\
\bf{H}_{lin} &= \frac{1}{\sqrt{2}}\left[ \bf{H}_+ \pm \bf{H}_-\right].
\end{align}
It is useful to think of this as analogous to representing linearly polarized light as a superposition of left and right circularly polarized.

\subsection{A note about the polarization}
\label{subsec:pol}

The polarization of the propagating field can be characterized at the ONF waist with polarimetric measurement of the scattered light. Because Rayleigh scattering is dipolar, the scattered field preserves the polarization of the incoming one. This allows us to measure the polarization of the guided field at each point along the nanofiber. We can control the polarization at the ONF waist by modifying the input polarization. For more details see \cite{Vetsch2012}.

The calculations in Sec. \ref{sec:fieldeqs} show that $E_z$ can be significant in an optical nanofiber \cite{LeKien2004OC}.  This is generally true for tightly-focused laser beams and waveguides in the strong-guiding regime.  The existence of longitudinal polarizations is intimately related to the first of Maxwell's equations (Gauss' law). While in everyday paraxial optics rays and transverse polarizations suffice to characterize optical phenomena, when there are significant gradients on the transverse field they have to be accompanied by a corresponding longitudinal component of the electric field. \cite{Lax1975} carries out a systematic (perturbational) approach starting with Gauss' law as a function of a small parameter $\lambda/w$ where $\lambda$ is the wavelength of the light and $w$ the characteristic transverse width of the field: the transverse size of the beam. As $w$ decreases the ray optics approximation with transverse fields fails and longitudinal components appear. This is a serious issue when focusing laser beams to small sizes and their polarization properties have been studied and measured in \cite{Erikson1994}.

Gauss' law ($\nabla \cdot {\bf E} =  0$) establishes that a focused light beam of wavelength $\lambda$ and angular frequency $\omega$ that propagates along $\pm z$ with a slowly-varying amplitude has a longitudinal component
\begin{equation}
\label{eq:chirality}
E_{z} = \pm \frac{i}{k} \nabla \cdot {\bf E}_{\perp}
\end{equation}
where the $\pm$ corresponds to the two possible directions of propagation and ${\bf E}_{{\perp}}$ is the transverse field.  The presence of $i$ in the expression for $E_z$ means that $E_z$ is $\pm\pi/2$ out of phase with the transverse components, depending on the propagation direction.  The existence of this longitudinal field and resulting transverse component of elliptical polarization is at the source of the chirality observed in nanophotonic systems as stated in \cite{Lodahl2016}.

\subsection{Higher-order modes}

As seen in Fig.~\ref{fig:neff}, when $V \leq 2.405$, the nanofiber supports only the two $\mathit{HE}_{11}$ degenerate modes. For a typical fused silica fiber, this requires $a\lesssim0.35\lambda$.  Nanofibers supporting higher-order modes (HOMs) provide additional degrees of freedom to enable complex evanescent field profiles for interaction with the surrounding medium.  A few works have explored the use for HOM in nanofibers for atom trapping~\cite{Sague2008,Phelan2013}, interactions with atoms~\cite{Kumar2015}, and high-resolution fiber profilometry~\cite{Hoffman2015,Fatemi2017}.

\begin{figure}
\centering
\includegraphics[width=0.75\textwidth]{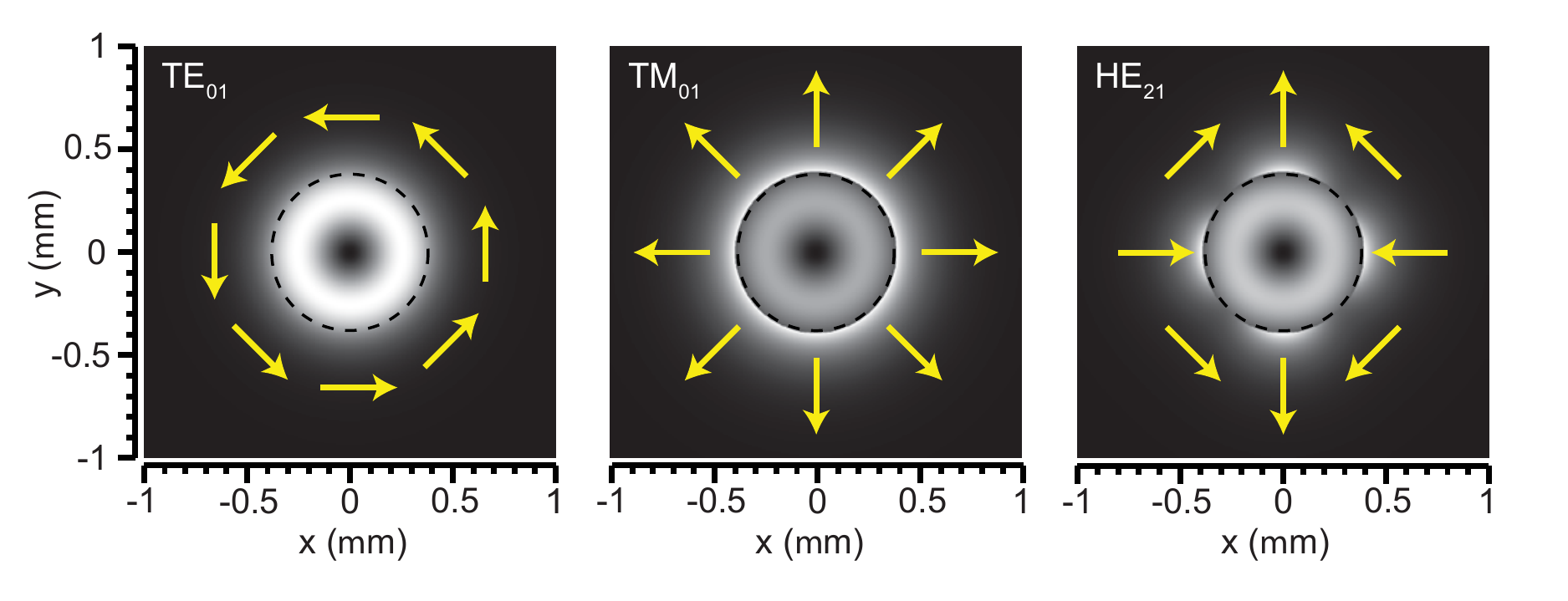}
\caption{\label{fig:transverse_modes}Transverse intensity profiles of modes in the $LP_{11}$ family, along with the $\phi$-dependent local polarization direction qualitatively indicated by the yellow arrows.  The ONF outline of the $a=390$ nm ONF is indicated by the dashed black line.}
\end{figure}

The treatment above provides modal solutions for these higher orders when $a$ increases.  Referring to Fig.~\ref{fig:neff}, the next grouping or family of modes is comprised of the $\mathit{TM}_{01}$, $\mathit{TE}_{01}$, and two degenerate $\mathit{HE}_{21}$ modes, with transverse profiles inside the fiber shown in Fig.~\ref{fig:transverse_modes} for an ONF with $a=390$~nm and $\lambda=795$~nm; in the weak-guiding limit, these are in the $\mathit{LP}_{11}$ basis.  For $\mathit{TM}_{0m}$ and $\mathit{TE}_{0m}$, only 3 of the 6 components of $\bf E$ and $\bf H$ are nonzero.  In particular, the $\mathit{TE}_{01}$ mode is purely azimuthally-polarized, having $E_z = E_r = 0$ everywhere; while the $\mathit{TM}_{01}$ mode is quasi-radially-polarized, with $E_\phi=0$.  These two modes are cylindrically symmetric and achieve cutoff at the same $V=$2.405.  The degenerate $\mathit{HE}_{21}$ modes have azimuthally-varying polarization, with transverse field components that alternate between radial and tangential polarization, and mode cutoff near $V=2.8$. These are labeled $\mathit{HE}_{21}^e$ and $\mathit{HE}_{21}^o$ with orthogonal transverse polarizations at every point. Direct observation of a nanofiber radius reaching cutoff condition during the tapering process is shown in ~\cite{Frawley2012a,Ravets2013}.

To control the HOM composition on the nanofiber waist, the field at the input of the standard optical fiber (Fig.~\ref{fig:nanofiber}) must also be controlled.  At the fiber input, the waveguide is in the weakly guiding regime, described by the LP-basis, with negligible longitudinal field and near unity overlap with superposition of the free-space propagation Hermite-Gauss (HG) or Laguerre-Gauss (LG) modes.  By tailoring an incident Gaussian beam with spatial light modulators~\cite{Frawley2012a} or appropriate phase plates and mode conversion~\cite{Ravets2013,Fatemi2011}, we can efficiently and selectively excite and control specific modes (or their superpositions) in the nanofiber waist ~\cite{Hoffman2015}.  Diagnostics of light propagation within the nanofiber using Rayleigh scattering~\cite{Hoffman2015} or near-field detection~\cite{Fatemi2017} have shown that mode discrimination is possible even in the subwavelength region of the waist.

\subsection{Light propagation in nanofibers}

As indicated at the beginning of this section, the nanofiber system consists of three distinct propagation regions:  The fiber input, made of standard optical fiber in the weakly-guiding regime; the nanofiber waist in the strongly-guiding regime; and the tapered region connecting both.  Computational analysis of light propagation from the input to the waist can be achieved with finite-difference time-domain (FDTD) methods.  The analytical treatment above is sufficient for the two-layer input and waist regions, having only two $n_i$, but within the taper region, the effect of finite cladding radius becomes strong and calculations must consider the third medium (usually air or vacuum). 

\begin{figure}
\centering
\includegraphics[width=0.75\textwidth]{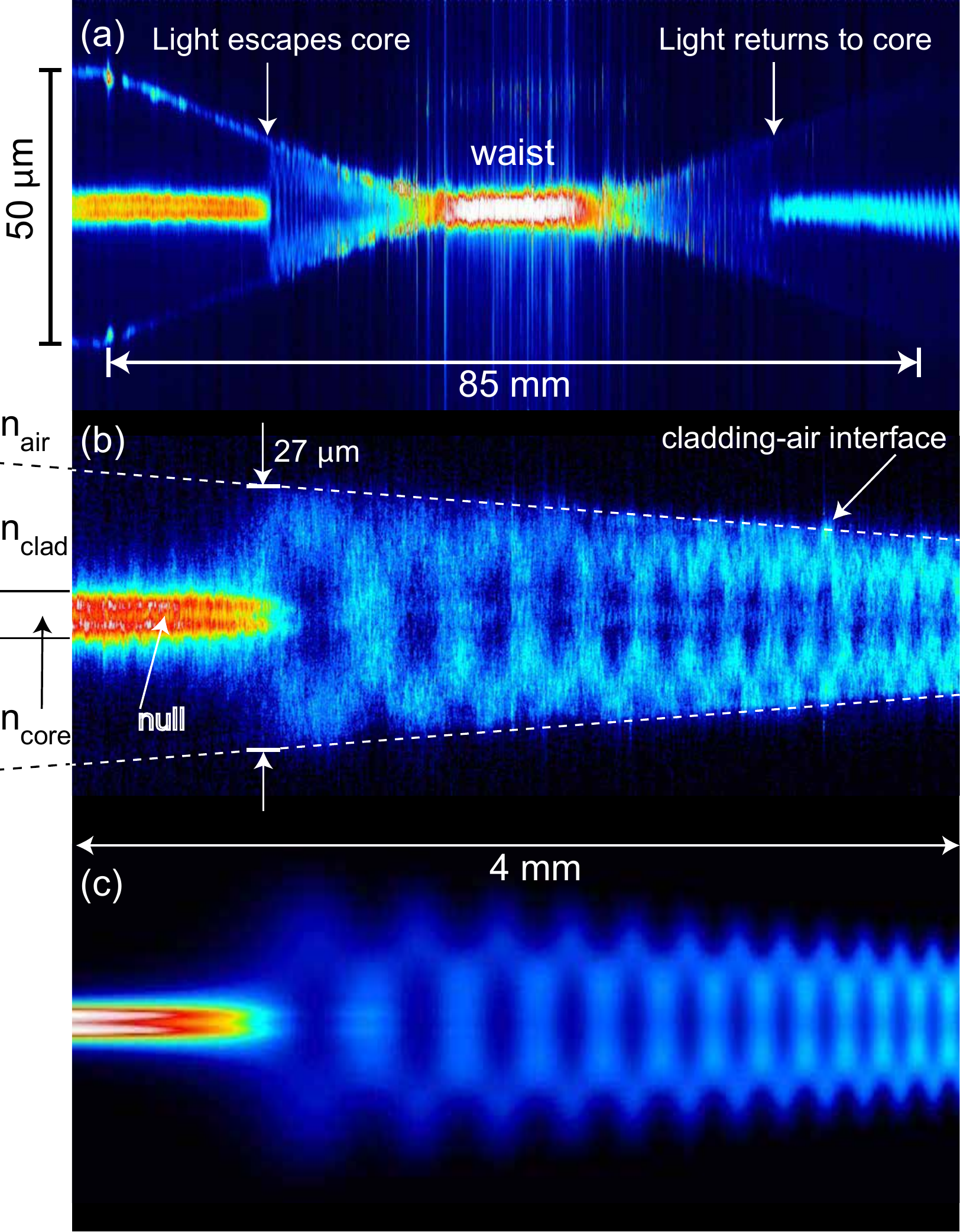}
\caption{\label{fig:Rayleigh}{Propagation of the $\mathit{LP}_{11}$ mode family through an ONF by imaging Rayleigh scattered light.  (a) shows the entire tapered fiber waveguide.  Most notable is the transition from core-cladding guidance to cladding air guidance, magnified in (b).  The transition is sudden, occurring over a radius change of only $\approx$20 nm. The intensity null observed in the core is due to the ``donut''-like intensity profile of this mode family (see Fig.~\ref{fig:transverse_modes}). (c) Simulated propagation using FIMMPROP.  The excellent agreement between simulation and experiment, and high dependence of spatial intensity beating on radius, enables a high-precision measure of fiber radius. Reprinted from Fig. 4 with
permission from  Hoffman, J. E., Fatemi, F. K., Beadie, G., Rolston, S. L., Orozco, L. A., 2015. Rayleigh scattering in
an optical nanofiber as a probe of higher-order mode propagation. Optica 2, 416. Copyright (2015) by the Optical Society}}
\end{figure}

The evolution of the light propagation within the nanofiber from core-cladding guidance to cladding-air guidance can be measured by imaging Rayleigh elastic scattering~\cite{Hoffman2015}, shown in Fig.~\ref{fig:Rayleigh}.  The figure shows propagation of the $LP_{01}$ mode family through the waveguide.  For the Fibercore SM1500 fiber used, the field leaves the core near $2a=27~\mu$m and enters the cladding volume.  The sudden large change from guiding with small $V$-number ($V\approx 2$) to large $V$-number ($V\approx 100$) represents a mode volume change of more than 1000.  It is this region that places the most stringent requirements on adiabatic propagation, as described in analytical treatments~\cite{Frawley2012a,Ravets2013a,Nagai2014}.

With multiple allowed modes, beating due to different modal propagation constants is easily detected.  Because the fields in cylindrically-symmetric waveguides are exactly solvable, and because the $\beta_i$ are strongly dependent on the nanofiber radius, the measured beat frequencies can be used as a nondestructive, \textit{in situ} method for evaluating the fiber radius with sub-angstrom radius sensitivity~\cite{Fatemi2017}.  Near-field methods based on whispering gallery modes \cite{Sumetsky2010, Sumetsky2006} or effects on transmitted light \cite{Madsen2016} can also provide sub-nanometer radius precision.
\newpage
  
\section{Fabrication and characterization}
\label{sec:fabrication}

Experiments in quantum optics and quantum information with atom traps around ONFs can benefit from high-transmission nanofibers. This reduces laser power requirements and unwanted stray light from both scattering and non-adiabatic mode excitation.
The review by \cite{Ward2014} analyzes a variety of ONF fabrication techniques. 

Common techniques for optical nanofiber production make use of micro-furnaces, fusion splicers, chemical etchants, and CO$_2$ lasers \cite{Ding2010, Lambelet1998, Kbashi2012, Yokota1997, Dimmick1999, Ward2006, Sorensen2014,Ward2014}. 
Chemical etching generally produces lower transmission than other heat and pull methods but offers more variety over the shape of the taper and the length of the waist. 
A CO$_2$ laser produces high-transmission optical nanofibers but the final diameter is limited by the power and focus of the laser.
Our work \cite{Hoffman2014a} follows closely that of~\cite{Warken2007b} and we summarize it in this section.

Our pulling technique is based on an existing methodology~\cite{Brambilla2010}.
It requires two pulling motors and a stationary oxyhydrogen heat source. 
This flame brushing method allows us to reliably produce ONFs with controllable taper geometries and uniform waists.  
With our setup the waist can vary in length from 1 to 100 mm, and we can achieve radii as small as 150 nanometers \cite{Bilodeau88, Birks1992, Warken2007,Warken2007b, Garcia-Fernandez2011}.   
Rather than sweep the flame back and forth over the fiber, we keep the flame stationary while sweeping the motors; this action reduces air currents, which could lead to nonuniformities on the fiber waist.
This approach is applicable to other pulling techniques as well, so there would be no need to scan a heat source.   

We focus on the critical pre-pull steps necessary to achieve an ultrahigh transmission, before handling the known environmental effects \cite{Fujiwara2011}.
Following the protocols described in this section we have produced fibers with 99.95\% transmission when launching the fundamental mode \cite{Hoffman2014a}. This high transmission corresponds to a loss of 2.6 $\times\,10^{-5}$~dB/mm for the fundamental mode \cite{Brambilla2010, Stiebeiner2010}, with controllable taper geometries and long fiber waists. 
We have also launched HOMs~\cite{Ravets2013a, Ravets2013} and achieved transmissions of greater than 97\% for the first family of excited modes. This level of transmission requires a thorough optimization of the pulling algorithm and fiber cleaning procedure. However, we can routinely achieve transmissions above 99 \% with root mean square (rms) fractional uniformities on the waist region smaller than 0.02. 

\subsection{The fiber-pulling apparatus}

The entire pulling apparatus is inside a nominally specified  ISO Class 100 cleanroom.  
The apparatus consists  of a heat source that brings the glass to a temperature greater than its softening point (1585$^{\circ}$ C for fused silica) and two motors that pull the fiber from both ends. 
We use two computer-controlled motors, mounted to a precision-ground granite slab.
The granite slab serves two purposes: its weight suppresses vibrations and it provides a flat surface. 
The weight of the granite slab
damps the recoil from the fiber motors as they change direction at the end of every pull step.  
Without a flat surface the motors will not work to specification, leading to distortion of the ONF during the pull: the pitch or yaw of the motor can vary the distance between the fiber and the flame, changing the effective size of the flame and pulling the fiber in unintended directions (negating any pre-pull alignment).  
The granite is mounted on an optical table at three points so that surface imperfections of the optical table do not distort the granite slab.

We attach flexure stages with a fiber clamp  to the moving platforms of the motors.  
We position the v-grooves of the fiber clamp on each stage at the minimum separation allowed by the parameters of a given pull.
Separating the fiber clamps at the closest allowed distance minimizes the fiber sag during the pull, which can result in the ONF breaking during the pulling process.  
The v-grooves of the fiber clamps on the left and right fiber motors must be aligned within a few micrometers to achieve a high transmission. 
We align the v-grooves  using  micrometers, attached to the flexure stages to allow for three axis translation assisted by an $in\ situ$  optical microscope.
The optical microscope is also used to check the pre-pull cleanliness of the fiber and for post-pull characterization. 
We use an oxyhydrogen flame to heat the fibers, in a stoichiometric mixture of hydrogen and oxygen to ensure that water vapor is the only byproduct. 
Stainless steel gas lines introduce the hydrogen and oxygen through flow meters.  
The gases mix in a tee and and pass through a high-quality, 3-nm filter. 
Finally the hydrogen-oxygen mixture exits through a custom-made nozzle with 29 holes of 200 $\mu$m diameter in a 1$\times$2 mm$^2$ area. 
The nozzle serves as a flame arrestor, while still allowing for the gas flow to be in the laminar regime.
The nozzle is clamped to a computer-controlled motorized stage that translates the flame to about 0.5 mm in front of the fiber for the duration of the pulling process. If any fiber buffer remains or dust lands on the fiber at any time the transmission will degrade.  

\subsection{Algorithm}\label{sec:alg} 

To choose the parameters for a pull we developed an algorithm\footnote{The program is available at the Digital Repository of the University of Maryland (DRUM) at http://hdl.handle.net/1903/15069.}, based on~\cite{Warken2007} and \cite{Warken2007b}, that calculates the trajectories of the motors needed to produce a fiber with the desired final radius, length of uniform waist, and taper geometry. 
The tapers are formed by a series of small fiber sections that are well-approximated by lines, allowing us to form a linear taper with a given angle down to a radius of 6 $\mu$m.
This connects to an exponential section that smoothly reduces and connects to a uniform section with a submicron radius, typically 250 nm. 
The slope of the linear taper section generally varies between 0.3 and 5 mrad. 
The algorithm divides the pull into steps defined by their pulling velocity and the traveling length of the flame.
It recursively calculates the parameters, starting from the desired radius, until reaching the initial one. 

\subsection{Cleaning and alignment procedure}
\label{sec:cleaningprocedure}
Obtaining high transmission through an ONF requires careful attention to the pre-pull cleanliness of the fiber.  
If any particulate remains from the fiber buffer or if dust arrives on the fiber before being introduced to the flame, the particulate will burn and greatly diminish the final transmission, see~\cite{Hoffman2014a}.  
Residue from solvents can also decrease transmission.  

Our cleaning procedure starts by mechanically removing the protective plastic buffer to expose the glass of the fiber to the flame. 
Then we use isopropyl alcohol on lens tissue to remove larger particulates. 
A few wipes of acetone
are then applied with class 10a cleanroom wipes, in order to dissolve smaller remnants of the buffer. 
A final cleaning with methanol using class 10a cleanroom wipes removes residues left from the previous solvents. 
We carefully lay the fiber into the grooves of the fiber clamps on the pulling apparatus and image the entire length of cleaned fiber using the {\it in situ} optical system. 
If there is any visible dust,  particulate, or evaporate, within the 2-$\mu$m resolution of the optical microscope, we repeat the cleaning procedure over.  
If the fiber is clean, we proceed to align it.  

We begin the alignment procedure by properly tensioning the fiber. Moving the pulling motors apart in 200-$\mu$m increments until the fiber slides through produces a uniform tension. 
We align the sections of fiber directly next to the left clamp and right clamp equidistant from the $in\ situ$ optical microscope and at the same height. There are more critical details about untensioning the fiber before starting the pull that the interested reader can find in \cite{Hoffman2014a}. The fiber alignment is on the order of micrometers over a length of centimeters.

\subsection{Characterization and microscopy validation}\label{sec:res}

We validate the accuracy of our algorithm expected fiber profile using both an $in\ situ$ optical microscope and a scanning electron microscope (SEM). The measurements of the radius with SEM agree with our targeted ONF radius to better than 5\%, while optical measurements across the whole fiber including the taper region give rms fractional variations around 2\%. We can also test the profile with exquisite (nanometric) sensitivity when using the beating spatial frequencies among higher order modes as seen through Rayleigh scattering \cite{Hoffman2015} and by evanescent coupling to a second fiber (sub-angstrom) \cite{Fatemi2017}. 

\subsection{Transmission}\label{sec:transmission}
\begin{figure}
\centering
\includegraphics[width=0.9\textwidth]{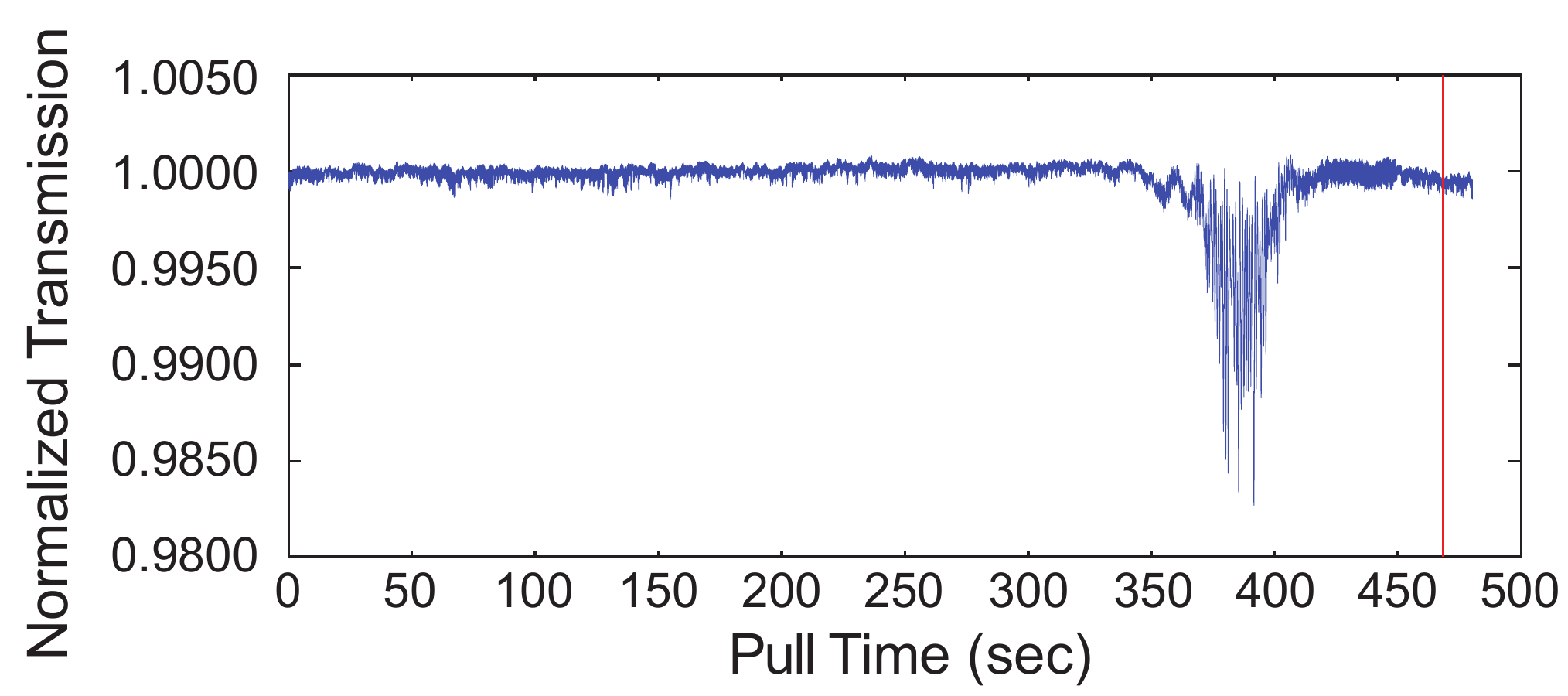}
\caption{The normalized transmission as a function of time during the pull through an ONF with an angle of 2~mrad to a radius of 6~$\mu$m and exponential profile to a final waist radius of 250~nm.  
The length of the waist is 5~mm.  
The fiber has a final transmission of 99.95~$\pm0.02$~\% or equivalently a loss of 2.6~$\,\times \,$~10$^{-5}$~dB/mm. The red line is the point where we evaluate the final transmission.
Reprinted with permission from Hoffman, J. E., Ravets, S., Grover, J. A., Solano, P., Kordell, P. R., Wong-Campos, J. D., Orozco, L. A.,
Rolston, S. L., Jun. 2014. Ultrahigh transmission optical nanobers. AIP Adv. 4 (6), 067124.
 Open access journal ~\cite{Hoffman2014a}.}
\label{fig:transmission}
\end{figure}
During the pull we monitor the transmission of a few nW of light through the fiber using a laser of 780-nm wavelength. 
We record data for the duration of the pull on a digital oscilloscope in high-resolution mode (16-bit digitization).   
We normalize the output signal to the laser power drift throughout the pull.
Fig.~\ref{fig:transmission} shows the transmission as a function of time during the pull for an ONF with a 2-mrad-angle taper to a radius of 6 $\mu$m and exponential profile to reach a final waist radius of 250 nm, with a fiber waist length of 5 mm.  
It achieves a transmission of 99.95~$\pm$~0.02~\%, corresponding to a loss of 2.6~$\times\,10^{-5}$~dB/mm when taken over the entire stretch.  

Using the numerical Maxwell's equations solver FIMMPROP, we calculate the expected transmission through a fiber with the same profile as in our pulls.  
We find the expected transmission to be consistent with our experimental result that measures the transmission through the entire ONF.  
Furthermore, when launching the next family of modes ($LP_{11}$) through the fiber the FIMMPROP simulations are well-matched to the achieved transmissions in~\cite{Ravets2013a}. 

\subsection{Spectrogram analysis}\label{sec:spec}

\begin{figure}
\centering
\includegraphics[width=0.85\textwidth]{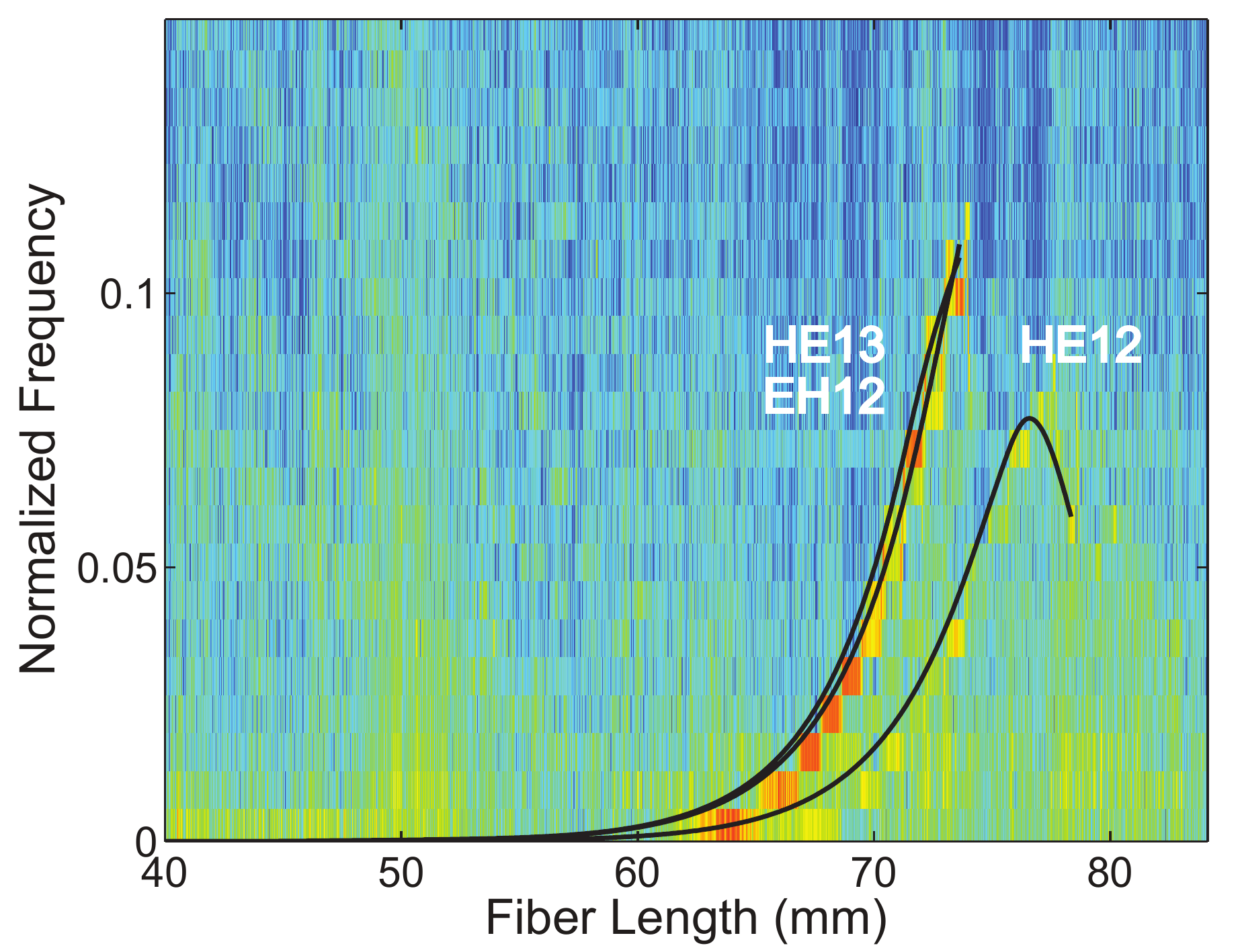}
\caption{Spectrogram of the transmission data from Fig.~\ref{fig:transmission} as a function of the stretch of the fiber, illustrating the
chirp of the beating frequency and the abrupt end of the beating. The colormap corresponds to the power spectral density. 
The  curves are calculations of the higher-order mode excitations of the same symmetry as the $HE_{11}$ fundamental mode: $EH_{11}$, $HE_{12}$, and $HE_{13}$. Reprinted Fig. 12 with
permission from Ravets, S., Hoffman, J. E., Kordell, P. R., Wong-Campos, J. D., Rolston, S. L., Orozco, L. A., 2013
Intermodal energy transfer in a tapered optical fiber: optimizing transmission. JOSA A 30, 2361, Copyright (2013) Optical Society~\cite{Ravets2013a}.
\label{fig:spec}}
\end {figure}

We analyze the quality of the ONF using a spectrogram, a short-time Fourier transform also sometimes referred to as the Gabor transform, of the transmission data during the pull.  
The spectrogram allows us to extract the modal change in the ONF during the pull.   
Each curve in the spectrogram corresponds to the evolution of the spatial beat frequency between the fundamental mode and excited modes propagating in the fiber, while the contrast corresponds to the energy transferred from the fundamental mode to the HOMs.  
We use simulations to identify all modes that are excited during the pull. \cite{Ravets2013a} has a detailed description and the full theoretical background.

Figure~\ref{fig:spec} is a spectrogram of the transmission data from Fig.~\ref{fig:transmission}.  For a successful 2-mrad pull with SM800 fiber we expect to observe few HOM excitations. If modes asymmetric to the fundamental are excited, we know the cylindrical symmetry of the fiber was broken during the pulling process \cite{Ravets2013a}, which can aid in identifying and fixing the error in the pulling apparatus.
We have seen that such coupling to asymmetric modes can occur for multiple-angle fiber pulls~\cite{Ravets2013} and uncleaned fibers \cite{Hoffman2014a}.

\subsection{ONF radius extraction}
\label{sec:radius}

We have developed a nondestructive tool to accurately and precisely measure the ONF radius and characterize the propagation of HOMs.  These properties are crucial for optimizing both fundamental and applied uses of ONFs. The ONF radius, $a$, governs the propagation of the HOMs through~Eq.~\ref{eq:V} and the coupling between a nearby atom (trapped or free) to the allowed nanofiber modes (see \cite{LeKien2005}, Secs.~\ref{sec:around}, and ~\ref{sec:trapping}).  As stated in Sec.~\ref{sec:modes} these modes have distinct effective refractive indices that depend strongly on the radius. 
 
Extraction of beating frequencies from Rayleigh scattering has the limitations of far-field imaging \cite{Hoffman2015,Szczurowski2011}. Contact techniques between an ONF and a probe microfiber avoid these limitations. They have been used to measure the local radius variations of an ONF with sub-angstrom/sub-nanometer precision by propagating light through the probe microfiber and observing the mode spectrum of a whispering gallery or a composite photonic crystal cavity~\cite{Birks2000, Sumetsky2010,Semenova2015,Keloth2015}. 
Other contact techniques rely on changes in the amplitude of the transmitted light, becoming sensitive to polarization and van der Waals forces, as in the recent work by~\cite{Madsen2016}, but do not directly sample the modal decomposition of the local field.

\begin{figure}
\centering
\includegraphics[width=0.7\textwidth]{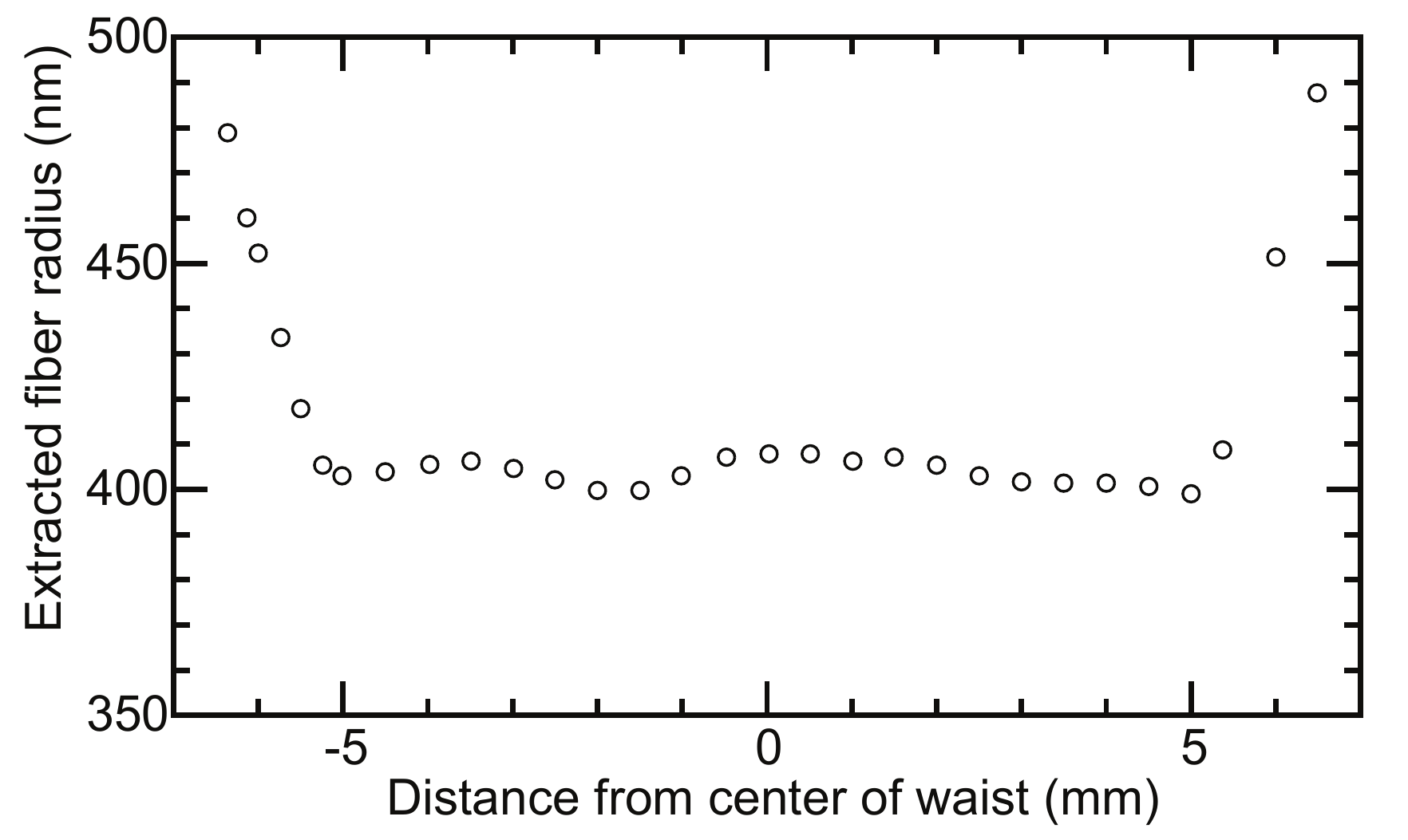}
\caption{\label{fig:radius}Radius extraction of a 10-mm nanofiber using the evanescent coupling \cite{Fatemi2017}. The radius comes from the $\mathit{HE}_{11}$:$\mathit{TM}_{01}$ beat frequency, using FFT windows every 600 microns. The error bars are smaller than the circles.}
\end{figure}

We measure the beat lengths between propagating nanofiber modes in the near field over the ONF length through evanescent coupling to a probe microfiber in \cite{Fatemi2017}. This near-field approach provides high ($\approx$1$~\mu$m) longitudinal resolution by taking the position dependent spectrogram of the coupled signal. The spatial frequencies map the mean of the ONF radius to within 0.4 Angstroms over 600-$\mu$m measurement windows along the ONF taper and waist. 
This technique has allowed us to measure how the nominal 380 nm radius waist changes over its 10 mm length finding it to be better than $\pm 5$ nm (the measured radius is closer to 400 nm), as Fig.~\ref{fig:radius} shows with a resolution smaller than the size of the data points (0.4 Angstroms). Such control on the fabrication process is beneficial for cold atom experiments in the vicinity of a nanofiber as detailed in sections \ref{sec:around}, and \ref{sec:trapping}.

\newpage
  
\section{Atoms around the nanofiber}
\label{sec:around}
This section presents studies of unconfined atoms and their interactions with ONFs. 
 Some experiments investigate the atom-surface interactions, others  use an ONF as a passive probe to measure properties of atomic clouds of atoms, while others perform linear and nonlinear spectroscopy using the nanofiber for light collection and/or directly probing the atoms.
 
The experiments in this section show the versatility of ONFs, and how they are making inroads on a wide variety of fields \cite{Brambilla2010}. We foresee many other uses, such as studies of molecules and molecular formation, and other forms of spectroscopy.

\subsection{Atom-surface interactions}
The interaction of atoms with a dielectric nanofiber has been studied theoretically in a number of papers~\cite{LeKien2005,LeKien2006,Klimov2004,Russell2009,Minogin2010,Frawley2012}.
They demonstrate that the nanofiber modifies the boundary conditions and density of modes, affecting decay properties of the atom.
These modifications can be measured in the frequency domain, through fluorescence and absorption spectroscopy, by observing lineshape asymmetries, or in the time domain through direct measurement of the spontaneous emission rate.

The van der Waals (vdW) and Casimir-Polder (CP) potentials will produce a red-shifted lineshape.
This has been measured in two experiments using cesium atoms.
The first observes the fluorescence spectrum, collected through the ONF, with a long red tail attributed to the vdW interaction~\cite{Nayak2007}.
They use the distance-dependence of the vdW potential to relate spectroscopic features to atoms passing at particular distances from the fiber surface.

\begin{figure}
\centering
\includegraphics[width=0.50\textwidth]{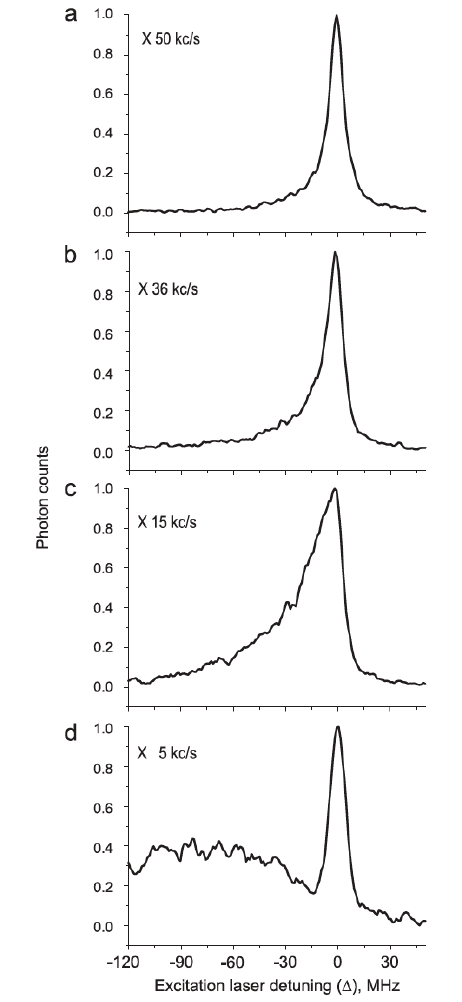}
\caption{Measured fluorescence excitation spectra.
The spectra (a), (b), (c) and (d) are measured one after another consecutively and the measurement time for each spectrum is around 6.5 minutes.
The scaling factors are shown in the figure.
The spectra are background corrected. Reprinted Fig. 2 with
permission from ,Nayak, K., Das, M., Le Kien, F., Hakuta, K., 2012. Spectroscopy of near-surface atoms using an optical
nanober. Opt. Commun. 285, 4698. Copyright (2012) Elsevier~\cite{Nayak2012a}.}
\label{fig:nayak}
\end{figure}

\cite{Nayak2012a} employ a similar technique with the addition of a violet laser to control the surface conditions of atoms adsorbed to the nanofiber.
After turning off the violet laser, they measure sequential spectra over time as more and more atoms adhere to the fiber, as displayed in Fig.~\ref{fig:nayak}.
The long tail on the red side of the spectrum, due to vdW interactions, becomes visible in Fig.~\ref{fig:nayak} (c) and (d).

\cite{Sague2007} instead use a low-power, through-fiber probe laser to study atom-surface interactions in a cloud of cesium atoms.
Because of the evanescent decay of the field on the fiber waist, the probe excites only atoms close to the nanofiber, and only on the order of nanowatts of laser power are needed to saturate the atoms.
At low powers, they see an asymmetry and broadening of the absorption profile due to light-induced dipole forces, the vdW potential, and emission enhancement.
As they increase the probe power, they observe a saturation of the linewidth, which is unexpected given power-broadening.
This is attributed to a reduced atomic density near the fiber because of the probe beam, which lowers the absorbance and narrows the line.
Simulations using atomic trajectories agree closely with their data in both the low- and high-power regimes.

Our group has also measured the transmission spectra of cold $^{87}$Rb atoms close to an ONF. We produce near-surface (<50 nm), thermally-desorbed atoms with a far-detuned, variable-intensity laser beam coupled to the ONF.
The spectra show asymmetries on the red side of the
transmission spectrum due to the van der Waals interaction of the atoms with the ONF~\cite{Patterson2016}.

\subsection{Atom-cloud characteristics}
Fluorescence from cold atom clouds is a key tool for determining properties of those clouds, e.g. temperature or atom number.
Typically the emitted photons are collected using free-space optics.
Another possible path is to employ an ONF to collect the photons emitted into the guided mode to discern atom-cloud characteristics.

Several groups have used nanofibers immersed in cold atom clouds in multiple experiments to both replicate and improve upon these standard techniques. \cite{Morrissey2009} uses the nanofiber to measure several characteristics of a magneto-optical trap (MOT): cloud shape, cloud size, atom loading rate, and atom number. 
In another experiment, \cite{Russell2012} measure sub-Doppler temperatures of an atom cloud by dithering the magnetic trapping field to extract the trap frequency.
The method is simultaneously corroborated with the standard free-space collection method. Finally, ~\cite{Russell2013} perform release and recapture with an ONF to measure atomic temperature, as well as diagnose optical misalignment in their MOT setups.

\begin{figure}
\centering
\includegraphics[width=0.75\textwidth]{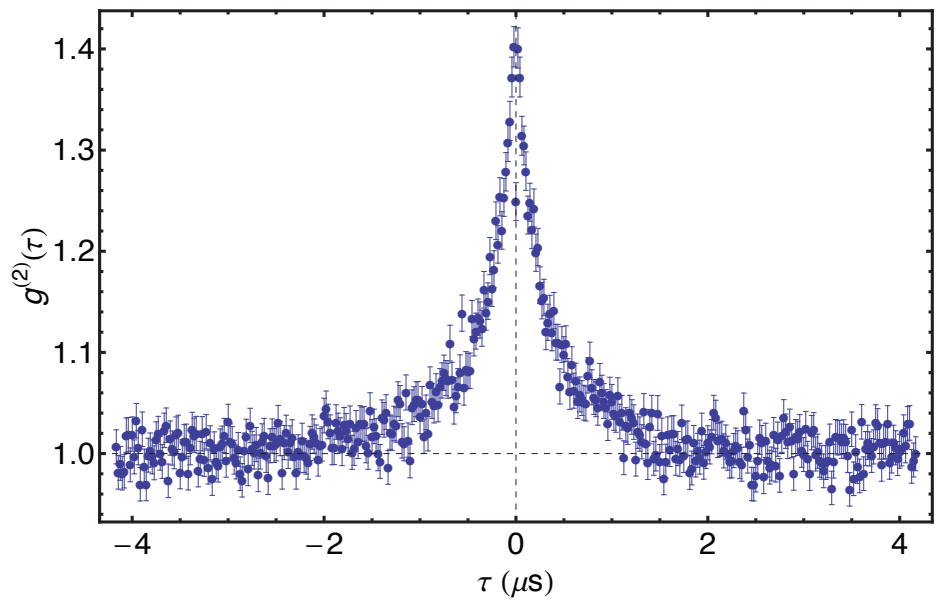}
\caption{\label{fig:grover}
Second-order correlation function $g^{(2)}_{\tau}$ as a function of delay time $\tau$ for a cold atom cloud that still shows an antibunching feature at $\tau=0$. 
}.
\end{figure}

Our photon-correlation measurements of \cite{Grover2015,Grover2015a} of a cloud of atoms around an ONF  allow us to
extract the atomic-cloud temperature from the correlations, by the knowledge of the evanescent field and the coupling of the moving atom fluorescence to the nanofiber mode. We validate the method against temperature measurements from time-of-flight images of the expanding atomic cloud. 
We measure the temporal width of the intensity-intensity correlation function due to atomic transit time, as presented in Fig.~\ref{fig:grover}, and use it to determine the most probable atomic velocity, hence the temperature. 
The results are further validated with atomic trajectory simulations.

\subsection{Linear and nonlinear spectroscopy}
The small mode profile of light propagating along the nanofiber waist can result in high intensities for even modest laser power.
As demonstrated in~\cite{Sague2007}, only nanowatts of resonant probe laser power are enough to saturate atoms near the fiber waist.
This makes nanofibers a nice platform for studying nonlinear optical effects that require high intensities.
Various low-power, nonlinear optical experiments have been conducted with nanofibers immersed in warm atomic vapors.

Using powers lower than 105 nW, ~\cite{Hendrickson2010} observe two-photon absorption in room temperature atoms around a nanofiber.
They attribute the neither Gaussian nor Lorentzian profiles of the two-photon absorption peaks to the transit time of atoms passing through the nanofiber mode.
In a deviation from typical alkali-atom experiments, \cite{Pittman2013} measure nonlinear transmission and low-power saturation through a nanofiber surrounded by metastable xenon atoms.
This is a promising avenue for for future nonlinear optics experiments since the inert xenon atoms will not accumulate on the fiber surface.
\cite{Jones2014} make a direct comparison between saturation effects in free-space versus through an ONF .
They present an empirical nonlinear absorption model that accurately fits the ONF data but fails when applied to the free-space data.

Another experiment measured the Autler-Townes splitting, in $^{87}$Rb, generating up-converted photons at 420 nm~\cite{Kumar2015b}.

\cite{Das2010} use quantum correlations to measure the fluorescence emission spectrum of a few strongly driven atoms around an ONF. 
Crucial to their success is the fact that they collect light on the nanofiber, then perform a heterodyne mixing before measuring the intensity-intensity correlation function, on which they perform a fast Fourier Transform to obtain the fluorescence spectrum. 
They observe that the fluorescence spectrum of the cold atoms near the nanofiber shows negligible effects of the atom-surface interaction, meaning they are far from the surface and the coupling $\beta$ is small, and agrees well with the Mollow triplet spectrum of free-space atoms at high excitation intensity.

\newpage
      
\section{Trapping atoms around an optical nanofiber}
\label{sec:trapping}
The evanescent field of the guided modes of an ONF provides an intensity gradient suitable for generating an optical dipole trap \cite{Grimm2000}. All current realizations use two lasers in the fundamental $\mathit{HE}_{11}$ guided mode \cite{Vetsch2010}. A laser red-detuned from resonance creates an attractive potential, while a blue-detuned one creates a repulsive potential preventing the atoms from sticking to the nanofiber surface. The characteristic decay length of both evanescent fields is proportional to their wavelengths (see Eq. \ref{eq:evanescent}), producing a longer (shorter) range attractive (repulsive) potential for the red- (blue-) detuned beam. By adjusting their relative intensities it is possible to find a configuration where there is a minimum of the dipole potential away from the ONF surface, achieving trap depths of a few hundreds of microkelvins. A longitudinal confinement of the trap atoms can be implemented by counter propagating two red detuned beams, creating a standing wave. Stable traps can be created for different trapping beam polarizations and intensities, generating different azimuthal confinements.

The gradient of the transversal component of the guided field is always on the order of $2\pi/\lambda$, creating a non-negligible longitudinal component (as explained in Sec. \ref{subsec:pol}), which is generally not the case for trapping beams propagating in free space, unless they are tightly focused. The phase relation between the transversal and the longitudinal component of the guided field creates elliptically polarized light, introducing significant vector light shifts in the trapped atoms \cite{LeKien2004}. Differential light shifts induced among atomic sub-levels can be suppressed with a proper choice of trapping wavelengths and polarizations \cite{Lacroute2012} or they can be used as a tool for atomic state preparation \cite{Albrecht2016}.

These types of trapping schemes have shown to be a promising platform for integrating laser-cooled atomic ensembles with optical fibers, opening a route towards applications in the context of quantum information and quantum technologies.

\subsection{An atomic dipole trap around an ONF}

\begin{figure}
\centering
\includegraphics[width=0.65\textwidth]{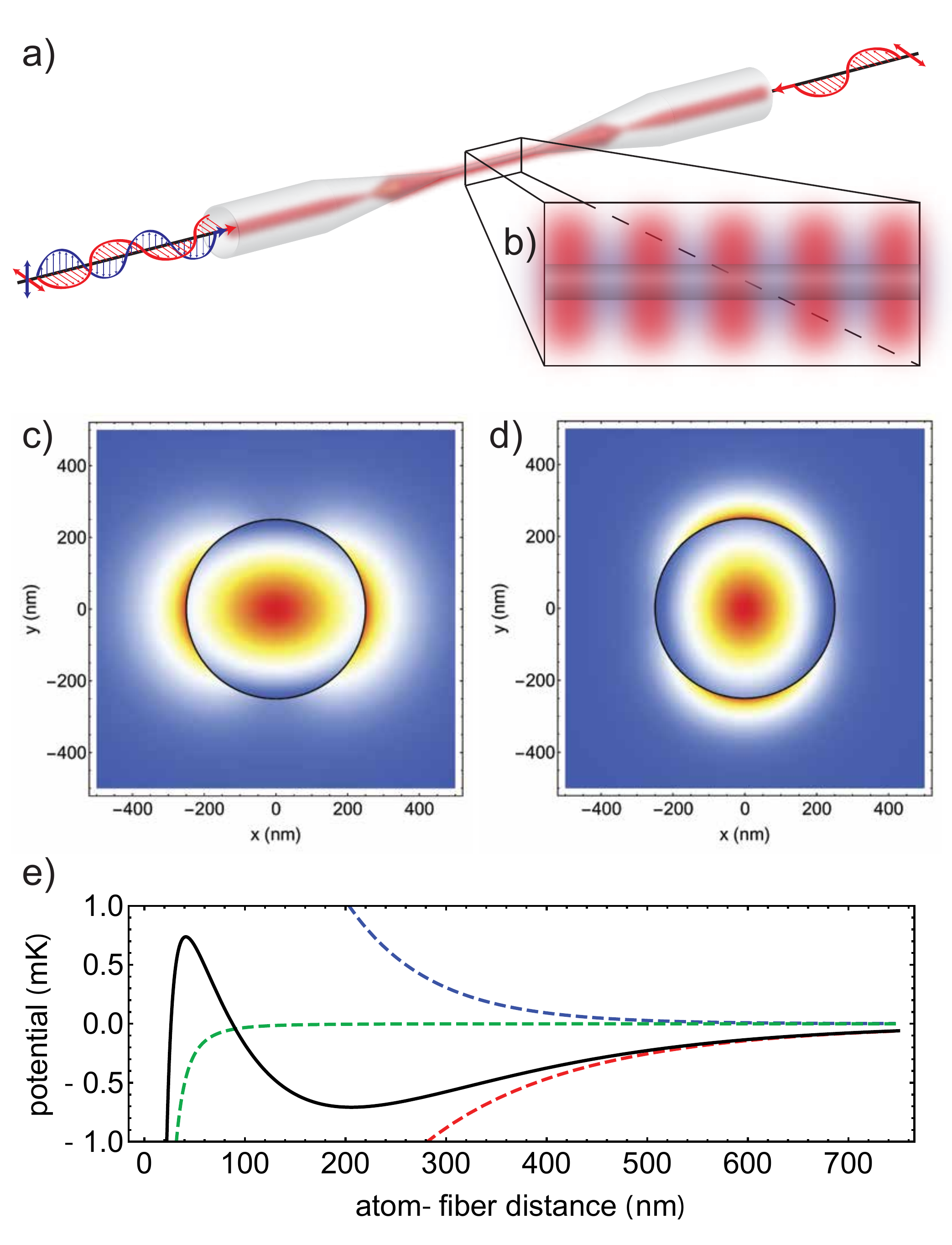}
\caption{\label{fig:potential}(a) Schematic of an ONF with counter-propagating 1064-nm beams and an orthogonally-polarized 750-nm beam. (b) Illustration of the intensity of the fields at the ONF waist with lattice formed by 1064-nm beams. (c) Intensity plot of quasilinearly-polarized, 1064-nm light in an ONF with diameter 500 nm. The color scale indicates increasing intensity from blue to red. (d) Intensity profile of vertically-polarized 750-nm light through the same ONF. (e) Total trapping potential (black solid) for a 500-nm diameter ONF with contributions from 3 mW in each 1064-nm beam (red dashed), 6.5 mW of 750-nm power (blue dotted), and van der Waals (green dashed and dotted). The potentials are calculated from only the scalar shifts. (Reprinted from Fig. 1.4 with
permission from  Hoffman, J. E., 2014. Optical nanofiber fabrication and analysis towards coupling atoms to superconducting
qubits. Ph.D. thesis, University of Maryland College Park. Copyright (2014) J. E. Hoffman~\cite{Hoffman2014}).}
\end{figure}

We now summarize how our own ONF trap operates \cite{Lee2014,Solano2017c} (see Fig. \ref{fig:potential}).
The ONF has a waist diameter of $500$ nm and length of 7 mm, with tapering regions of 28\,mm in length and 2 mrad angle. A MOT loaded from a background vapor of $\mathrm{^{87}Rb}$ produces a cloud of $\sim 10^8$ atoms. We overlap the cloud with the ONF waist using magnetic field shim coils and a ultra-high vacuum mechanical manipulator. Two orthogonal imaging systems ensure the overlap between the two of them. 
Atoms fall into the ONF trap (the red- and blue-detuned trap beams are on throughout the experiment) after 90 ms of increased MOT detuning  and a 1-ms-duration optical molasses stage, cooling the atoms to $\mathrm{\sim15}\,\mu K$.

An ONF trap requires light that is tuned to the red of the resonant frequency (with respect to the $^{87}$Rb D2 line) to provide an attractive potential and light tuned blue of resonance to prevent atoms from striking the ONF surface (see Fig.~\ref{fig:potential}). A 1064-nm wavelength laser beam provides the (red) attractive potential and a 750-nm wavelength beam the (blue) repulsive force. A potential minimum of a few hundred $\mu$K in depth is formed $\sim$200\,nm from the fiber surface, as calculated with a simple two level atom and only scalar shifts (see Fig.~\ref{fig:potential} (e)).  

Counter-propagating the red-detuned attractive beams, generating a standing wave, confines the atoms along the nanofiber. This configuration creates a one-dimensional lattice on each side of the ONF. The presence of a longitudinal component of the propagating electric field prevents us from having a standing wave with perfect contrast. The contrast varies for different wavelength and ONF radius, and in our case is $\sim$65\%. However, this is not an impediment for trapping the atoms in a lattice configuration.

The azimuthal confinement of the trapped atoms is achieved by using quasi-linearly polarized trapping beams in the $\mathit{HE}_{11}$ mode (see Fig.~\ref{fig:potential} (c and d)). There is no particular requirement for the relative polarization of the blue and red detuned beams. A trapping minimum can be generated for polarizations varying from parallel \cite{Goban2012} to perpendicular \cite{Vetsch2012}, provided the proper optical powers of the trapping beams are used. To verify the polarization of each beam on the nanofiber waist, we take polarization-sensitive measurements of the Rayleigh scattering of the propagating beams at the ONF waist.

\begin{figure}
\centering
\includegraphics[width=0.7\textwidth]{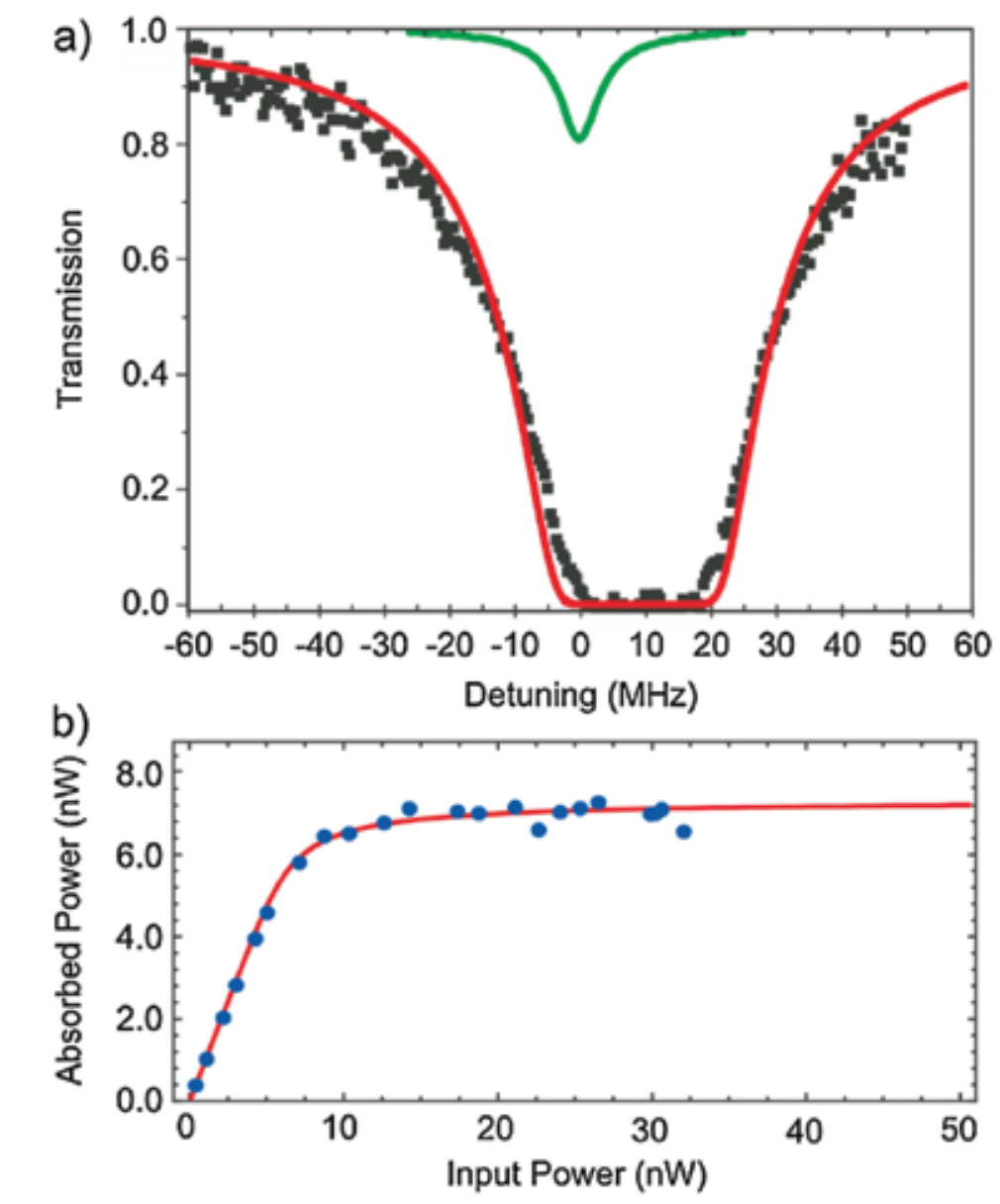}
\caption{\label{fig:Vetsch}(a) Transmission spectrum of the probe beam
through the nanofiber after loading the trap (black squares). For
reference, the spectrum of the MOT cloud (green line) is plotted.
The red line is a theoretical fit, see \cite{Vetsch2010}. (b) Saturation measurement yielding the number of trapped atoms. The
red line is a theoretical fit. (Reprinted Fig. 3 with permission from Vetsch, E., Reitz, D., Sagu{\'e}, G., Schmidt, R., Dawkins, S. T., Rauschenbeutel, A., 2010. Optical
interface created by laser-cooled atoms trapped in the evanescent eld surrounding an optical nanofiber.
Phys. Rev. Lett. 104 (20), 203603, with
Copyright (2010)
by the American Physical Society~\cite{Vetsch2010}).}
\end{figure}

Figure \ref{fig:Vetsch} shows results from the pioneering experiment by \cite{Vetsch2010}. The authors demostrate trapping and optical interfacing of neutral Cs atoms around an ONF. The atoms are localized in two one-dimensional optical lattice arrays about 200 nm above and below the nanofiber surface and can be efficiently interrogated with a resonant light field sent through the nanofiber. Their operation parameters allow only zero or one atoms on each independent well because of collisional blockade \cite{Schlosser2002}, creating on average a filling factor of 50\%. 

The number of trapped atoms in an ONF two-color dipole trap can be measured by propagating a probe beam through the nanofiber. Possible strategies are based on transmission spectroscopy and off-resonance dispersive measurements.

\subsection{Transmission spectroscopy}

Transmission spectroscopy is based on the Beer-Lambert-Bouguer law for optical absorption in a thick medium and the knowledge of the atomic spectral lineshape. By scanning the frequency of a weak near-resonance probe beam and measuring its transmission as a function of frequency, we can obtain the optical density of the sample \cite{Foot2005}. If the atom-light coupling strength is known, the OD can be translated to a number of trapped atoms. When implemented in an ONF the probe power is typically of the order pW, to avoid saturation of the atoms and power broadening. The transmitted photons are counted with avalanche photodiodes and appropriate electronics. A careful bandwidth filtering is necessary to keep the background from the blue detuned laser and what appears to be fluorescence of the fiber. Figure \ref{fig:Vetsch}(a) shows an example of this measurement. Similar results can be found in \cite{Goban2012}. 

Nonetheless, the number of trapped atoms cannot always be simply determined by a Lorentzian line shape of the probe beam absorption. This depends on the trapping potentials and the atomic species. Ref. \cite{Lee2014} shows that distinct asymmetries are observed in the spectrum due to the effects of the vector light shifts associated with the optical trapping fields. Although Rb and Cs are nominally atoms with  similar atomic structure, the light shifts can in fact be quite different, with larger effects in Rb than Cs, leading to a modified absorption profile. 

This method of measuring the number of trapped atoms is a rather destructive one. The absorbed photon transfers a momentum kick to the atoms. After repeated scattering events, the atoms heat up, and eventually leave the trap. 

\subsection{Dispersive measurements}

A less destructive atom-number measurement is to send an off-resonance probe beam through the ONF. The dispersive atom-light interaction creates an effective modification of the refractive index experienced by the propagating field, but minimally altering the atomic state. The tight mode confinement provides a large atom-light nonlinear interaction and efficient readout. The main features of this technique are explored theoretically in detail by \cite{Qi2016}, in the context of quantum non-demolition measurements. The dispersive atom-light interaction produces a phase shift in the propagating field, proportional to the number of atoms and the atom-light coupling strength. This phase shift can be read out with polarimetric \cite{Dawkins2011, Solano2017c} and interferometric measurements \cite{Beguin2014a}.

\begin{figure}
\centering
\includegraphics[width=0.65\textwidth]{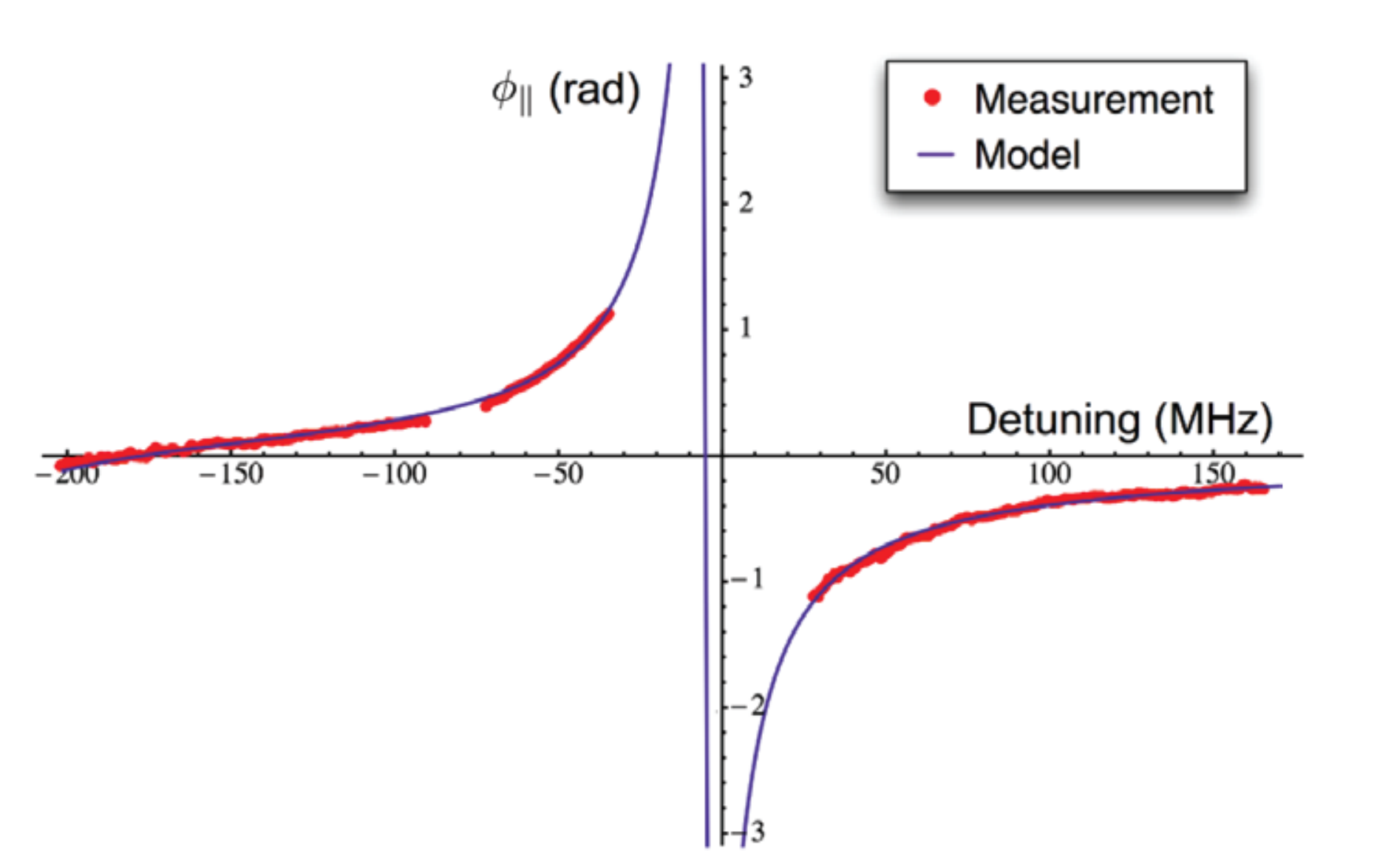}
\caption{\label{fig:Dawkins} Phase shift $\phi_{\parallel}$ of the eigenmode e$_{\parallel}$ induced by 1000 atoms and measured as a function of the detuning from the $F=4 \rightarrow F'= 5$ free-space transition frequency, $(\omega-\omega_5)/2\pi$, with 128 averages (red dots). The blue line is a fit taking into account the two nearest transitions $F=4 \rightarrow F'=4$ and $F=4 \rightarrow F'=5$. The former gives rise to the zero-crossing of the signal near-180 MHz. (Reprinted Fig. 3 with permission from Dawkins, S., Mitsch, R., Reitz, D., Vetsch, E., Rauschenbeutel, A., 2011. Dispersive Optical Interface
Based on Nanober-Trapped Atoms. Phys. Rev. Lett. 107, 243601, with
Copyright (2011)
by the American Physical Society.~\cite{Dawkins2011}).}
\end{figure}

The polarimetric measurement relies on the azimuthally asymmetric geometry of the trap. An ensemble of trapped atoms at each side of the ONF defines a preferential propagation axis for the evanescent field of an off-resonant quasi-linearly polarized probe beam. The atoms create an effective birefringence for the probe. In \cite{Dawkins2011}, 1000 cesium atoms are trapped around an ONF, producing a phase shift per atom of $\approx 1$ mrad at a detuning of 6 times the natural linewidth. Fig. \ref{fig:Dawkins} shows the measured phase shift for different probe detuning.

\cite{Beguin2014a} realize an interferometric measurement of the probe beam. They send two probe beams detuned by the same frequency above and below resonance. This eliminates any possibility of modification of the potential landscape experienced by the trapped atoms. Because the sign of the phase shift depends on which side of the resonant frequency the probe is, both probe beams experience equal-magnitude but opposite-sign phase shifts. The differential phase shift between the two probes is measured using optical homodyne interferometry, allowing them to extract the number of trapped atoms.

\subsection{State-sensitive and state-insensitive traps}

When a two level atom interacts with an oscillating electric field, \textit{e.g.} a dipole trap, the two energy levels shift in opposite directions, an effect known as the ac-Stark shift. If the field is red-detuned from the atomic resonance, the ground state shifts down in energy and the excited state shifts up; the opposite case is true for a blue-detuned optical beam. When the shift only depends on the total amplitude of the electric field it is called scalar light shift, and when it depends on field polarization it is called vector light shift. In a dipole trap produced by circularly polarized light the polarization of the electric field couples to the components of the angular momentum of the real (multilevel) atom, shifting also the hyperfine sub-levels, similar to what happens to the atoms in the presence of a ``fictitious" external magnetic field. This is the case for ONF-based dipole traps, due to the significant longitudinal component of the propagating field that produces elliptically-polarized light along the nanofiber. The differential shifts of the atomic energy levels cause different electronic states to experience different trapping potentials. State-dependent potentials create difficulties for coherent control of atoms, because internal states will couple to the noisier center-of-mass motion. 

We can solve this issue using the multilevel structure of real atoms. In a multilevel-atom the scalar shift that the first excited atomic state experiences due to its coupling with the ground states can be counteracted by an opposite shift due to its coupling with further-excited states. The wavelength of light that creates the same shift for the ground and excited atomic states is called a magic wavelength. The idea of magic wavelengths, originally stated by \cite{Cho1997} in the context of a possible systematic effect in precision measurements, shows that in a dipole trap the ground and excited electronic states of an atom can experience the same trapping potential, permitting coherent control of electronic transitions independent of the atomic center-of-mass motion \cite{Ye2008, Arora2012}. On the other hand, a vector light shift is not present when using only linearly polarized light. In the case of an ONF dipole trap, this can be achieved by canceling the longitudinal component of the electric field by using equal intensities and polarization for the counter-propagating trapping beams. A frequency difference between blue-detuned counter-propagating beams creates a traveling wave that averages to produce a uniform repulsive potential. These methods for implementing a state-insensitive dipole trap in an ONF are proposed in \cite{LeKien2005b} and first implemented in \cite{Goban2012, Lacroute2012,Ding2012}, where they trap Cs atoms $215$ nm from the surface of a nanofiber, and suppress the differential scalar and vector light shifts by a factor of 250.

State-sensitive traps, which remove the degeneracy of the atomic hyperfine-sublevel, can be used for state preparation and interrogation of trapped atoms. For these applications, the vector light shift in ONF-based dipole trap can be particularly useful \cite{LeKien2013, Albrecht2016}. Applying additional real or fictitious magnetic fields, the state dependence of the trapping potential can be controlled, providing a means to probe and to manipulate the motional state of the atoms in the trap by driving transitions between Zeeman sub-levels. In particular, microwave sideband cooling can be implemented to bring the trapped atoms to the motional ground-state \cite{Albrecht2016}. Another possible application is to use the fictitious magnetic field induced by a nanofiber-guided light field in conjunction with an external magnetic bias field to create an effective trapping potential for atoms around the ONF \cite{Schneeweiss2014}.

\subsection{Ground-state coherence}

Despite the difficulties that state-dependent trapping potentials impose in the coherent control of arbitrary atomic states, ground states can keep coherences for times suitable for some quantum information applications. \cite{Reitz2013} study the ground state coherence properties of Cs atoms in a nanofiber-based two-color dipole trap. Using microwave radiation to coherently drive the clock transition, they record Ramsey fringes as well as spin echo signals and infer a reversible dephasing time of $T^{*}_{2}=0.6$ ms and an irreversible dephasing time of $T'_{2}=3.7$ ms. Both time constants are limited by the finite initial temperature of the atomic ensemble and the heating rate, respectively. 

\subsection{Reflectivity}

The regime of collective coupling with high OD provided by ONF-based dipole traps opens a door to the study of long-range interactions and, for example, the formation of so-called atomic mirrors~\cite{Chang2012} and the possible observation of self-crystallization~\cite{Chang2013, Griesser2013}. 

A high reflectivity due to an array of atoms requires the separation between the atoms to be commensurate with the resonance wavelength \cite{Deutsch1995}. In general, this condition is not satisfied for atoms trapped around an ONF. The counter-propagating red-detuned beams that provide the lattice configuration for trapping the atoms is usually far detuned from resonance, helping to avoid losses from spontaneous emission. This imposes an extra difficulty for the implementation of optical reflectors that has to be circumvented.

\begin{figure}
\centering
\includegraphics[width=0.75\textwidth]{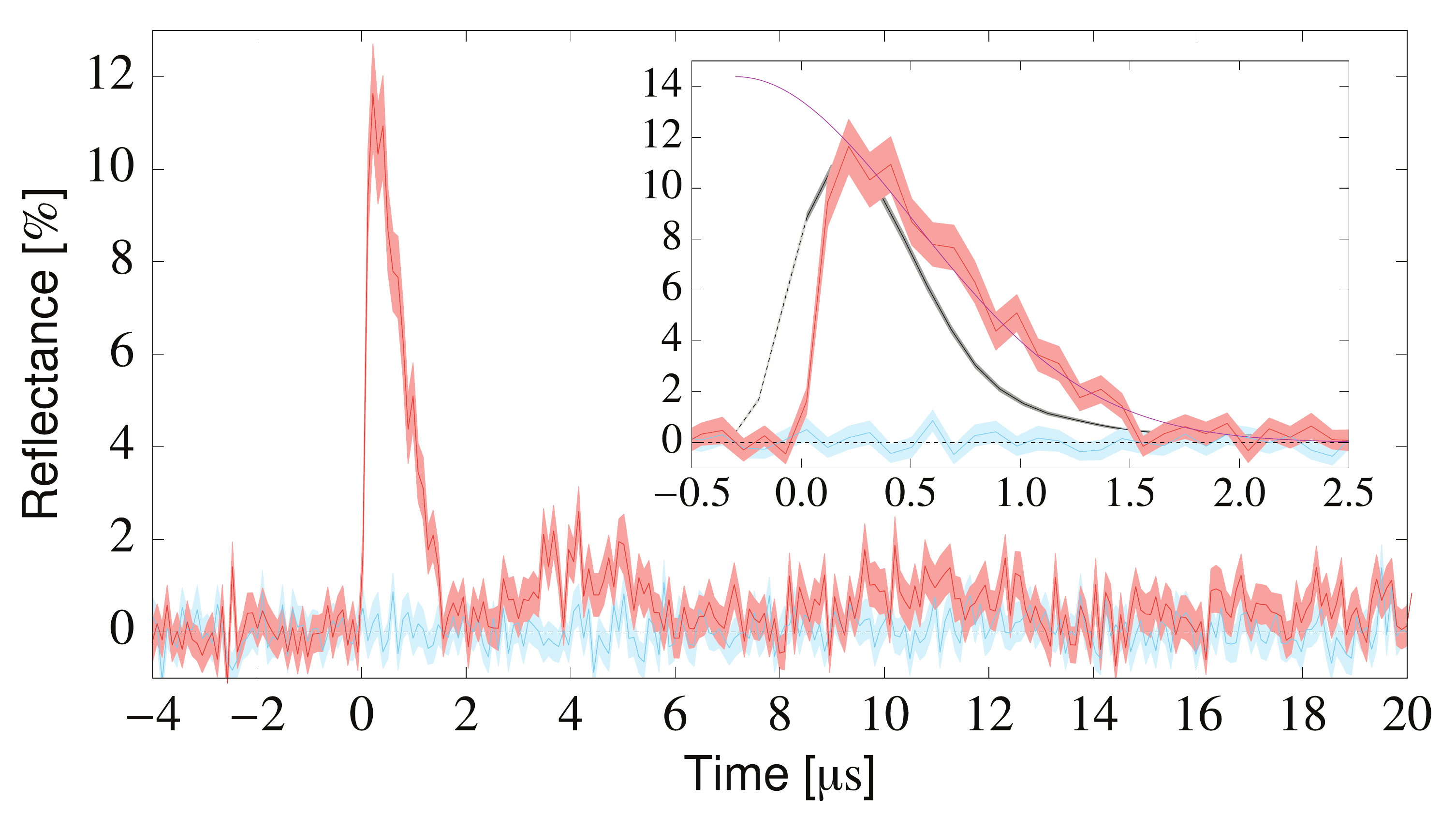}
\caption{\label{fig:Sorensen}Reflectance off the atomic strings within a 2.8 MHz
detection bandwidth (3 dB). The probe is tuned 8 MHz above
atomic resonance and turned on at t=0 with an independently
measured rise time of 80 ns. Blue curve: Unstructured atomic
ensemble, average of 250 experiments. Red curve: Structured
ensemble, average of 200 experiments. The shaded regions signify
the 1$\sigma$ uncertainty interval with contributions from statistical
averaging and a 5\% input probe power fluctuation during the
measurement. Inset: Zoom on the first reflection peak with a fit to a
Gaussian decay (purple) and the theoretical prediction (black) with
uncertainty band given by the statistical averaging. (Reprinted Fig. 2 with permission from S\o{}rensen, H. L., B\'eguin J.-B., Kluge, K. W., Iakoupov, I., S\o{}rensen, A. S., M\"uller, J. H., Polzik, E. S.,
Appel, J., 2016. Coherent backscattering of light o one-dimensional atomic strings. Phys. Rev. Lett.
117, 133604.
Copyright (2016)
by the American Physical Society.~\cite{Sorensen2016}).}
\end{figure}

Two groups have recently reported coherent backscattering of light (reflectivity) off trapped Cs atoms around a nanofiber. \cite{Sorensen2016}  shows  coherent Bragg scattering in a one-dimensional system realized by trapping atoms in the evanescent field of an ONF. They report nearly 12\% power
reflection of a probe light sent through the nanofiber interacting with only about 1000 cesium atoms, an enhancement of two orders of magnitude
compared to the reflection from randomly positioned atoms. This is achieved by optically pumping the atoms to the appropriate state with a standing wave that imprints the necessary periodicity. The main mechanism that lowers the reflectivity seems to be the dephasing from the oscillations of the atoms in each individual trap that start at different places in the potentials. Fig.~\ref{fig:Sorensen} shows the reflected intensity and its decay. There is a revival of the reflectivity about 4 $\mu$s after the structure pulse, showing a brief re-phasing of some of the pumped atoms oscillating in the trap.

\begin{figure}
\centering
\includegraphics[width=0.75\textwidth]{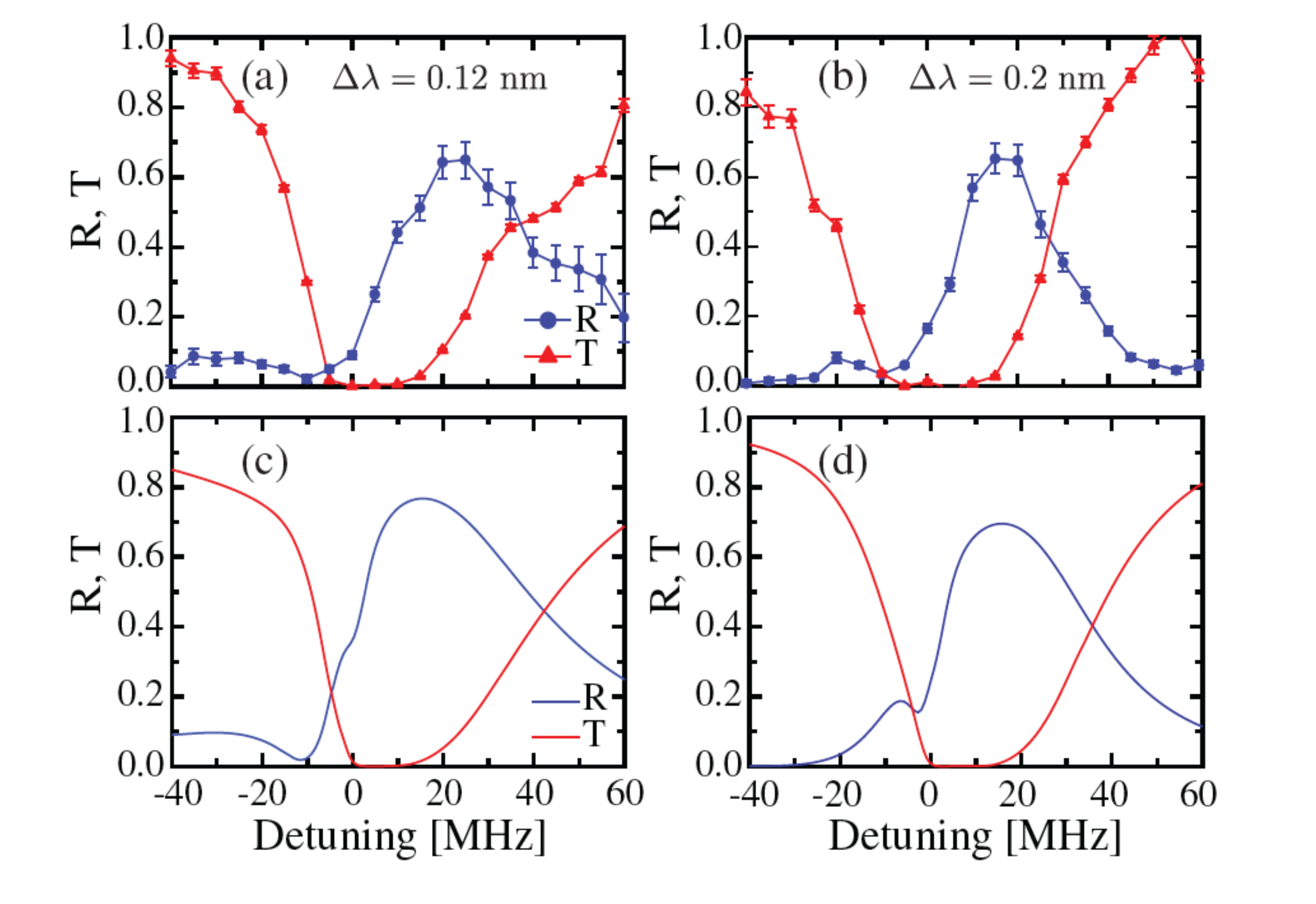}
\caption{\label{fig:Corzo}
Reflection and transmission spectra for a probe quasi-linearly
polarized along the y direction. (a) and (b) Experimental
results for $\Delta\lambda$=0.12 nm and $\Delta\lambda$=0.2 nm. (c) and (d) Simulated
spectra for N=2000 atoms, $\gamma_{\mbox{1D}}/\gamma_{0}=0.007$, and a filling factor
of the lattice sites is 0.3. The coupling value and filling factor
have been adjusted to fit the measured spectra reported in (a). 
(Reprinted Fig.3 with permission from Corzo, N. V., Gouraud, B., Chandra, A., Goban, A., Sheremet, A. S., Kupriyanov, D., Laurat, J., 2016.
Large bragg reflection from one-dimensional chains of trapped atoms near a nanoscale waveguide. Phys.
Rev. Lett. 117, 133603.
Copyright (2016)
by the American Physical Society.~\cite{Corzo2016}).}
\end{figure}

\cite{Corzo2016} also report Bragg reflection from atoms trapped on a nanofiber.  By using an optical lattice with a period nearly commensurate with the resonant wavelength, realized by a pair of close-to-resonance red-detuned counterpropagating beams, they observe a reflectance of up to 75\%. Each atom behaves as a partially reflecting mirror and an ordered chain of about 2000 atoms is sufficient to realize an efficient Bragg mirror. They report measurements of the reflection spectra
as a function of the lattice period and the probe polarization (See Fig.~\ref{fig:Corzo}).

\subsection{Trapping with higher-order modes}

Allowing higher-order modes to propagate through an ONF enables new optical trapping geometries. These modes have a transversal intensity profile and a field polarization richer than the fundamental mode (see Sec. \ref{sec:modes} and Fig. \ref{fig:transverse_modes}), facilitating modified transversal confinements of the atoms, usually further away from the ONF surface. Different modes also have different propagation constants, creating an interference pattern along the nanofiber as they propagate. This enables different possibilities for trapping patterns in the longitudinal direction \cite{Sague2008b} .

\cite{Phelan2013} investigate trapping geometries that combine the fundamental mode with one of the next lowest possible modes, namely the \textit{HE}$_{21}$ mode. Counter-propagating, red-detuned \textit{HE}$_{21}$ modes are combined with a blue-detuned \textit{HE}$_{11}$ fundamental mode to form a potential in the shape of four intertwined spirals. By changing the polarization from circular to linear in each of the two counter-propagating \textit{HE}$_{21}$ modes simultaneously, the 4-helix configuration can be transformed into a lattice configuration. Helical trapping geometries are also proposed for combinations of different polarizations of beams only in the fundamental \textit{HE}$_{11}$ mode \cite{Reitz2012}

The interaction between atoms and higher-order propagating modes can produce mode-conversion, the redistribution of power of an incoming field from one mode to another \cite{Kumar2015c}. The different propagation constant of the modes implies different momentum for each propagating mode. When a process of mode-conversion happens the differential change in momentum is compensated by a momentum transfer to the nearby atom. Based on this effect, recent papers have explored the possibility of using higher-order modes to control center-of-mass motion of trapped atoms \cite{ebongue2016} and trapped microparticles \cite{maimaiti2015} along the nanofiber.

\subsection{Possible heating mechanisms}

Atoms trapped around an ONF eventually leave the trap. The heating mechanisms that leads the atoms to escape have not been fully characterized. One contribution comes from elastic scattering events of the far-detuned trapping beams \cite{Grimm2000}, a well-known effect in dipole traps. A less familiar contribution comes from torsional modes of the ONF \cite{Wuttke2013}. The torsional modes couple to the polarization of all the guided fields. Since the trapping potential is sensitive to the polarization of the trapping beam, the torsional modes create a time-dependent perturbation of the potential. The frequencies of these modes are densely populated and they are close enough to the trapping frequencies ($\sim 200$ kHz) to consider parametric heating of the atoms as particularly strong mechanisms for losses. Torsional modes have been measured and characterized \cite{Wuttke2013}, and even optically excited \cite{Fatemi2016DAMOP,Fenton16}. However, it is necessary to suppress them or increase their frequencies to prevent direct and parametric heating of the atoms.

\newpage
     
\section{Quantum optics and quantum information with ONFs}
\label{sec:qo}
Section \ref{sec:introduction} has motivated the importance of ONFs for quantum optics and quantum information. They provide a unique platform for many reasons, one being that the presence of an ONF modifies the structure of the vacuum-electromagnetic field experienced by a nearby atom. This can modify atomic properties such as coherence or spontaneous emission rate, the latter attributed to the Purcell effect, \cite{Klimov2004,LeKien2005,LeKien2008a,LeKien2008c,Verhart2014}.  

Second, ONF guided modes create a preferential channel for the atomic radiation field. This is a virtually lossless channel that mediates interaction between distant atoms along the nanofiber, even when they are not trapped. When the interaction time is within the coherence of the system, they present a novel realization of many-body states in one-dimension with long-range interaction.

Third, atom-light interactions mediated by ONFs can be further increased by modifying the geometry of the platform. For example, through the introduction of Bragg mirrors in the optical fiber, the single atom cooperativity can reach the strong coupling regime.

Interfacing guided light with atomic ensembles has been actively pursued. Atoms around ONFs have been used as quantum information storage for memories and repeaters using the coherence of atoms similarly to EIT schemes.

\subsection{Geometric dependence of the atom light coupling}
We  start by  presenting calculations related to atom-light coupling (Sec.~\ref{sec:introduction}) relevant for quantum optics experiments. 
We look first to the emission enhancement coefficient, $\alpha$, of Eq.~\ref{eq:alpha}. The proximity of the nanofiber changes the spontaneous emission rate of the atom, and this change is reflected in this parameter~\cite{Klimov2004}. It will serve as a proxy by which we can maximize atom-light coupling through nanofiber and optical trap design.

\begin{figure}
\centering
\includegraphics[width=1.0\textwidth]{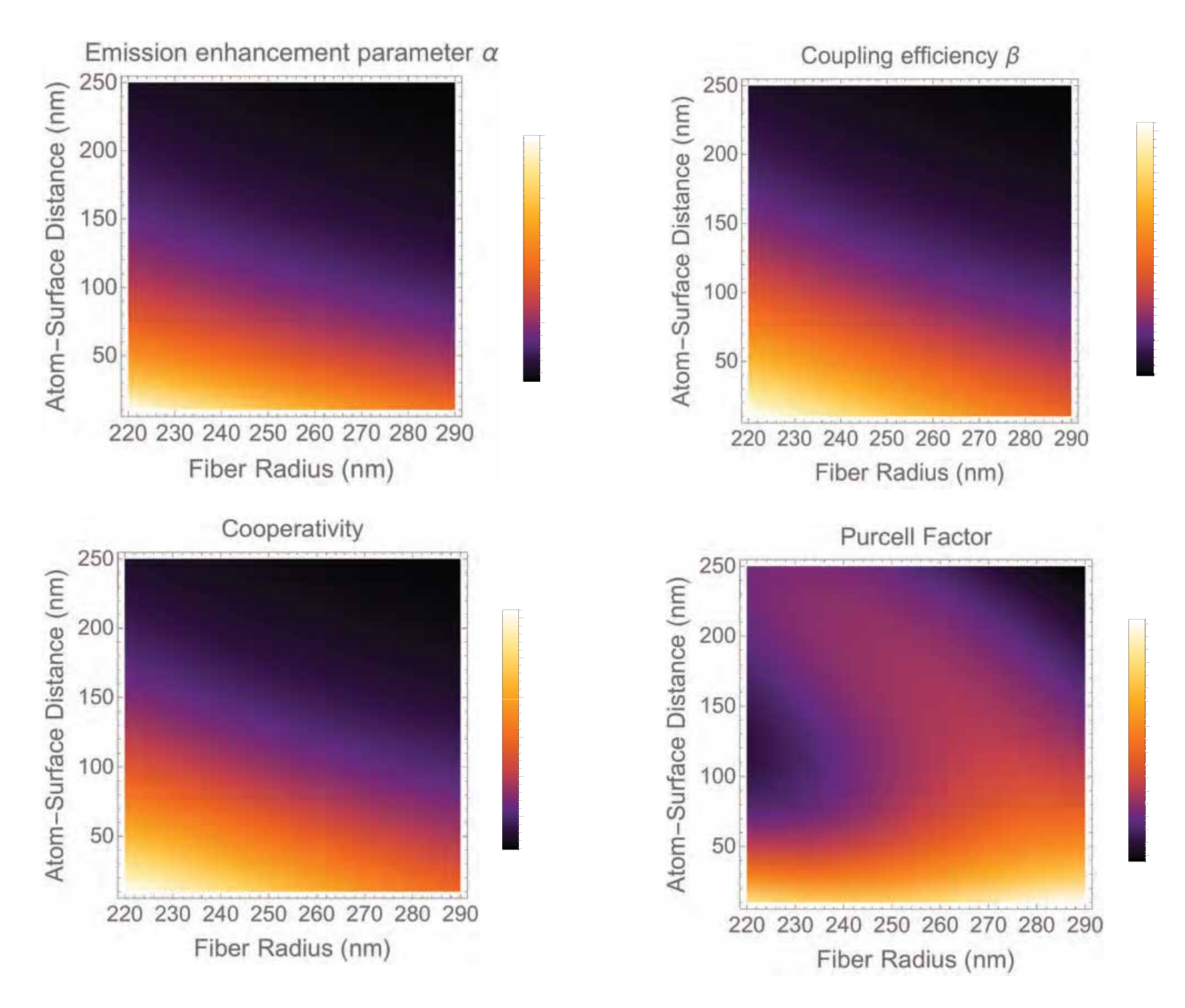}
\caption{Atom-mode coupling parameters as a function of nanofiber radius and atom-surface distance for our ONF configuration from \cite{Grover2015}. 
(a) Emission enhancement $\alpha$, (b) coupling efficiency, $\beta$  (c) single atom cooperativity $C_1$, and (d) Purcell factor $\alpha/\beta$. The Purcell factor, $C_{1}$, and $\beta$ depends strongly on the atomic dipole alignment relative to the ONF surface. Here we assume that the atomic dipole moment points in the same direction than the electric field of the circularly polarized fundamental mode of the ONF. 
}
\label{fig:alphabetaCPF}
\end{figure}

Figure~\ref{fig:alphabetaCPF} shows a calculation based on the mode structure in Sec~\ref{sec:modes} for a single Rb atom operating on the D2 line in the $\mathit{HE}_{11}$ evanescent mode of the ONF, following the work of~\cite{LeKien2005}.
The four relevant parameters - emission enhancement $\alpha$, coupling coefficient $\beta$, cooperativity $C_1$, and Purcell factor - are plotted as a function of fiber radius and atom-surface distance.

The emission enhancement can be up to 20\% larger than that of free-space near to the fiber surface according to Fig.~\ref{fig:alphabetaCPF}(a). As expected, it decreases to the free space value as the atom-fiber distance increases. For a typical nanofiber $\beta\ll 1$, it follows from Eq. \ref{eq:Cbeta} that $C_1\approx\beta$, as a simple comparison between of the plots in Fig.~\ref{fig:alphabetaCPF}(b) and (c) shows. Both the coupling efficiency and cooperativity display similar spatial dependence as the emission enhancement. Performing a global search we find that for a given mode there is an optimal ONF radius; reducing the radius decreases the coupling. The modification of the total spontaneous emission due to the presence of the nanofiber is given by the Purcell factor ($\alpha/\beta$), plotted in Fig.~\ref{fig:alphabetaCPF}(d). Its non-trivial dependence on the geometrical parameters comes from the atomic dipole alignment and is discussed in more detail in Sec.~\ref{subsec:PF}.

\subsection{Purcell effect of atoms around an ONF}
\label{subsec:PF}

The Purcell effect is the change of the atomic spontaneous emission due to the modification of the vacuum electromagnetic field in the presence of an object \cite{Purcell1946}. As explained in Sec. \ref{sec:coop}, it tells us about the properties of atoms coupled to a given structure of modes, in our case radiated (reservoir) and guided modes. When only the coupling into the preferential guided mode is known, the measurement of a Purcell factor helps obtain information about the coupling of the system to the reservoir. This provides us with a tool to characterize coherence properties of the system.

\begin{figure}
\centering
\includegraphics[width=0.75\textwidth]{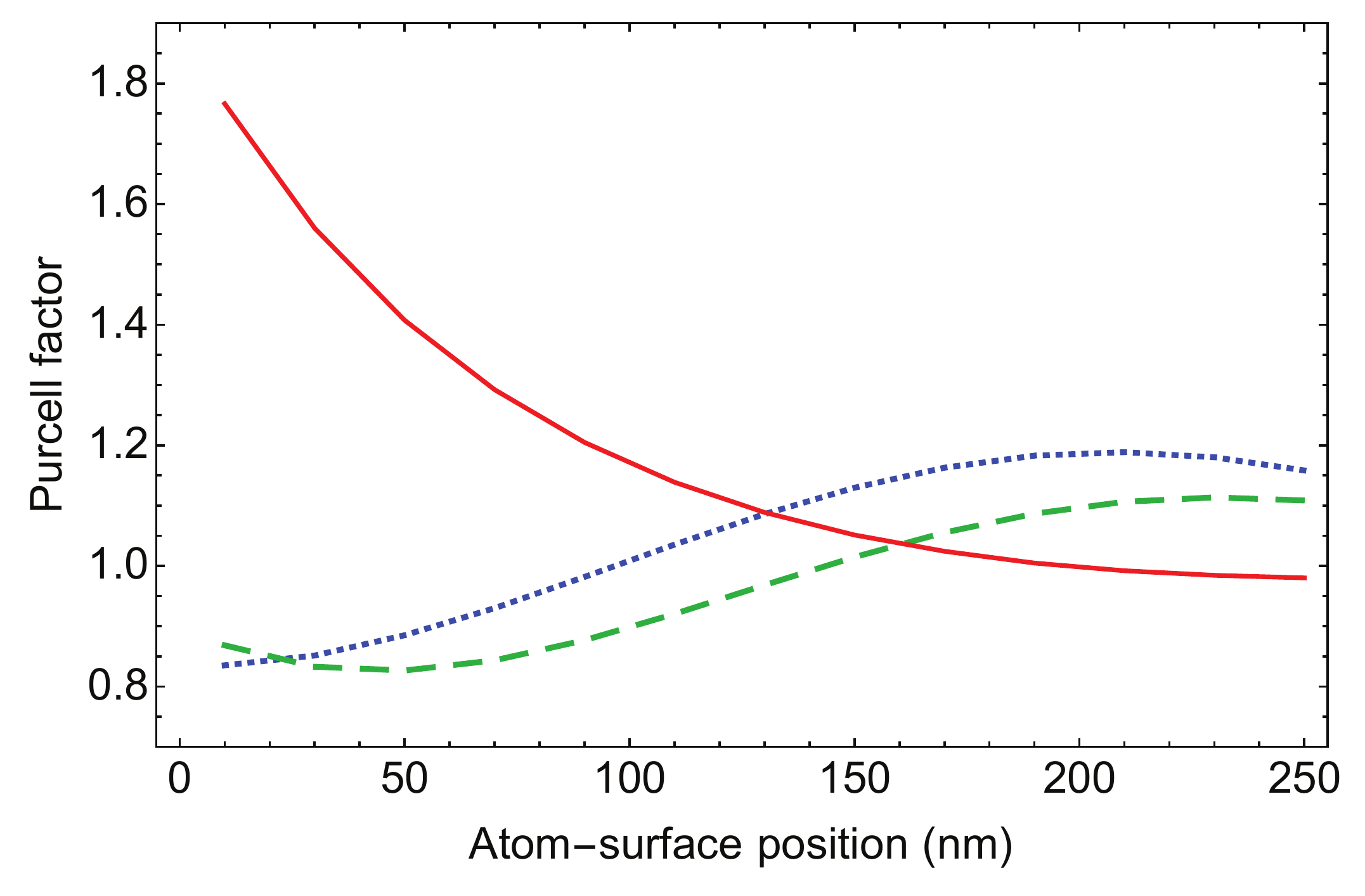}
\caption{\label{fig:PFs} Purcell factor of an ONF of 237-nm radius as a function of the atom-to-surface distance for an atomic dipole aligned perpendicular to (solid red), parallel to (dotted blue), and along (dashed green) the ONF. The curves are calculated from finite-difference time-domain calculations.}
\end{figure}

\cite{Solano2017b} presents measurements of the radiative lifetime changes in the vicinity of an ONF due to the presence of a dielectric surface and to atomic dipole alignment effects. A dielectric surface modifies the dipole moment of a proximal atom, in turn changing its rate of spontaneous emission in a manner analogous to the Purcell effect. The alignment of the induced atomic dipoles from the excitation relative to the ONF surface determines if there is an enhancement or inhibition of the atomic decay rate. The difference can be as large as $\sim$80\% enhancement to $\sim$20\% inhibition for our current ONF parameters. Fig. \ref{fig:PFs} shows this dependence, suggesting that for a particular alignment atomic dipoles could couple differently to the environment (particularly non-guided modes). We use a sample of free cold atoms around the nanofiber. In this configuration, it is possible to probe closer distances than with trapped atoms and still avoid the complications of the van der Waals interaction. We measure enhancements and inhibitions of the spontaneous emission rate of $\sim$10\% and $\sim$5\% respectively, depending on the alignment of the induced dipoles and averaged over the given atomic distribution. FDTD simulations are in agreement with the measurements.

\subsection{ONF-mediated coherent interactions}
\label{subsec:coherent}

\cite{Chang2012} look at the ONF platform and find a way to convert it into a cavity QED system with atomic mirrors where it is possible to reach strong coupling. See Sec. \ref{sec:trapping} for the recent results on the reflectivity of atoms trapped around an ONF. 

\cite{Griesser2013} study the collective off-resonant scattering of coherent light by a cold gas in a 1D configuration such as that of atoms trapped around a ONF. This scattering induces long-range interactions via interference of light scattered by different particles that can propagate through the ONF. They study the possible phase transitions of the atomic spatial order as a function of the external pump intensity which can show self-sustained crystallization, similar to \cite{Domokos2002}, but the parameter space is rich and there can coexist multiple ordered states with distinct appearances. Similar studies have been carried for slightly different nanophotonic platforms \cite{Chang2013, Holzmann2014,Eldredge2016}. For a detailed discussion about the role of long-range interaction in self-organized systems see \cite{Ritsch2013}.

Collective effects become readily visible for a strong enough coupling efficiency. By using optical pumping to positionally select the atoms such that their average distance is 30 nm from the ONF surface \cite{Solano2017a} reach an efficiency of $\beta\sim$0.13. This allows coherent interactions of atomic ensembles mediated by the nanofiber. Collective effects are evidenced by measuring the dependence of the atomic decay rate on the number of atoms, observing subradiant and superadiant behavior. In particular, superradiant decay times are observed for atoms separated by hundreds of resonant wavelengths, providing evidence of long-range coherent interactions.

\subsection{Antibunching with atoms around an ONF}
\label{subsec:Photon-counting}
Intensity-intensity correlation measurements are another class of experiments that can be applied to untrapped atoms around nanofibers.
Widely used throughout quantum optics, the intensity autocorrelation function, $g^{(2)}(\tau)$, measures correlations in the fluctuations of light intensity, e.g. the photon statistics, and can reveal both classical and quantum aspects of the light and its sources.

Using these methods, \cite{Nayak2008} observe the presence of a single atom emitting photons into the guided mode of the fiber.
A second paper enhances these results by varying the atom number in the vicinity of the nanofiber and by considering the direction of emission of the photons~\cite{Nayak2009}.
For photons emitted into the same direction, they observe a clear transition from antibunching to bunching at zero time delay as the number of atoms increases, as expected from the theory.
When measuring the correlations of photons emitted into opposite ends of the nanofiber, however, they find that antibunching persists regardless of atom number, attributable to a geometric effect depending on the phase of the excitation field.  An example of our work is on  Fig.`\ref{fig:grover}.

\subsection{EIT and optical memories}

Experiments have recently demonstrated the relevance of ONFs for quantum information and for general quantum optics. These experiments use the tight confinement of the field to induce non-linear interactions of atoms. The experiments use atoms that are  trapped, cold, or even at room temperature to observe EIT \cite{Fleischhauer2005} and exploit these properties together with well-established quantum memory protocols to store information and retrieve it \cite{Duan2001,Lvovsky2009}.

The theoretical treatment of \cite{LeKien2015} provides a framework to study the propagation of guided light along an array of three-level atoms in the vicinity of an optical nanofiber under the condition of EIT. They study different polarization arrangements and find that when the input guided light is quasilinearly polarized along the major principal axis, the transmissivity and the group delay of the transmitted field substantially depend on the propagation direction of the probe field. Under the Bragg resonance condition, an array of atoms prepared in an appropriate internal state can transmit guided light polarized along the major principal axis in one specific direction. They state the importance of the longitudinal polarization component for these effects.
 
An experiment with warm ($\approx373$~K) atoms by \cite {Spillane2008} reports the observation of low-light level optical interactions through an ONF. They show V-type EIT with continuous wave powers below 10 nW in the D1 manifold of $^{85}$Rb. They were sensitive to the transit time dephasing given the small mode cross section of the ONF.

Two recent experiments with Rb use the transitions between the $5S_{1/2} \rightarrow 5P_{3/2} \rightarrow 5D_{5/2}$ to establish a ladder configuration for EIT.  \cite{Jones2015}
use an optical nanofiber (300 nm diameter) suspended in a warm rubidium vapor with the natural abundance of isotopes. Their signal and control fields propagate through the nanofiber. Transit-time broadening, given the room temperature velocities is a significant EIT decoherence mechanism in this tightly confined waveguiding geometry. Their setup includes a way to keep the nanofiber warm to avoid deposits of Rb metal on the surface that prevent the propagation of light. They  observe significant EIT and controlled polarization rotation using control-field powers of only a few nanowatts in this warm-atom nanofiber system.

\cite{Kumar2015a} use cold $^{87}$Rb atoms at about 200 $\mu$K from a MOT to realize EIT also in the ladder configuration with the two fields (probe at 780 nm) for the $5S_{1/2} \rightarrow 5P_{3/2}$ transition  (coupling strong field at 556 nm) for the $5P_{3/2} \rightarrow 5D_{5/2}$. The fields propagate in opposite directions but overlap in the waist of the 350 nm diameter ONF. They are able to see the multilevel cascaded EIT. They observe and analyze the multipeak transparency spectra of the probe beam. They demonstrate all-optical-switching in the all-fiber system modulating the 80 nW coupling beam in the EIT region by observing the probe beam ($\approx 5$ pW) that inputs with a constant intensity and shows modulation at the output.

\begin{figure}
\centering
\includegraphics[width=0.65\textwidth]{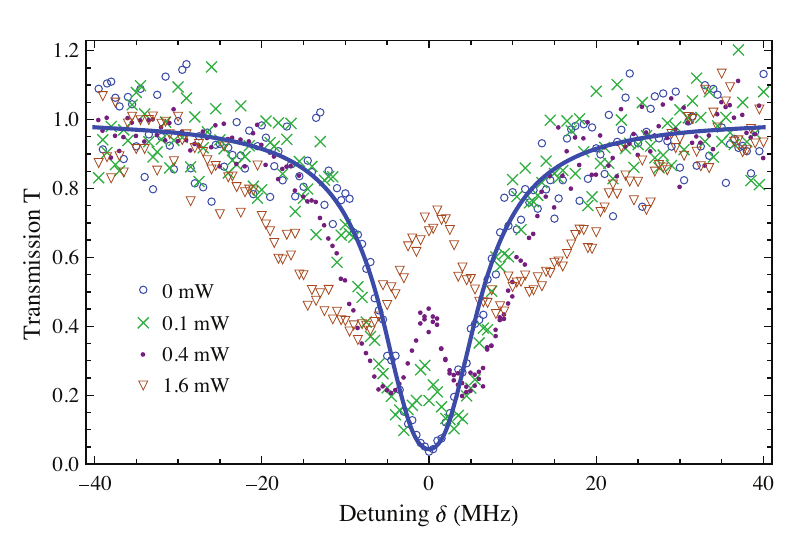}
\caption{\label{fig:Gouraud}
EIT for the guided light. The control is on resonance on the $|s\rangle\rightarrow|e\rangle$ transition while the signal is detuned by $\delta$ from the $|g\rangle\rightarrow|e\rangle$ resonance ($|s\rangle$, $|g\rangle$ and $|e\rangle$ are the two ground and the excited states respectively in $\Lambda$ configuration). Profiles are displayed as a function of $\delta$, for four values of the control power. (Reprinted Fig. 2 with permission from Gouraud, B., Maxein, D., Nicolas, A., Morin, O., Laurat, J., 2015. Demonstration of a memory for
tightly guided light in an optical nanober. Phys. Rev. Lett. 114, 180503, with
Copyright (2015)
by the American Physical Society.
of ~\cite{Gouraud2015}).}
\end{figure}

\cite{Gouraud2015} work with cold Cs atoms around their 400-nm-diameter ONF in a well-compensated magnetic environment. They use a $\Lambda$ system in the D2 line and involve the two hyperfine ground states. A narrow transparency window is observed for one of the transitions (signal), as Fig. \ref{fig:Gouraud} shows. The polarization control is important and they use Rayleigh scattering to ensure its properties (see \ref{subsec:pol}). The control field is a free-space beam propagating through the atomic cloud, while the signal from the upper hyperfine ground state ($F=4$) is coupled through the ONF.  They interact with about 2000 atoms and achieve coherent storage of a pulse of light with intensities low enough to be single photons. They obtain an overall efficiency of retrieval of $10 \pm 0.5$\% despite the limited delay available in the system from the EIT, the pulse cannot be contained entirely in the ensemble, and a leakage is observed before the control is switched off.  

\cite{Sayrin2015} use an ONF interfaced to an ensemble of trapped neutral
Cs atoms to store fiber-guided light.
They demonstrate EIT, slow light, and the storage of fiber-guided optical pulses in an ensemble of trapped cold atoms. They measure light group velocities of 50 m/s and store optical pulses at the single-photon level and retrieve them on demand in the fiber after 2 $\mu$s with an overall efficiency of $3.0\pm0.4$\%. They work on the D2 line of Cs with their probe between the low ($F=3, m_F=3$) hyperfine state detuned from the excited $6P_{3/2}, F'=4$ state, while they use as control the transition that starts on the $6S_{1/2}, F=4, m_F=4$ in the presence of a magnetic field perpendicular to the propagation direction of the light on the ONF. The control and probe beams copropagate through the ONF. The spectral transparency window can be as narrow as 30 kHz with a transparency of 60\%. They store and retrieve classical light pulses that contained fewer than one photon on average, and should be able to show  storage of quantum states with non-classical correlations in the future.
 
\subsection{Cavity QED with modified ONFs}
\label{sec:modifications}

One of the advantages of using ONFs for  atom-light interaction is the high single atom cooperativity achieved due to the high mode confinement. However, there are methods to further increase the single atom cooperativity as stated in Sec. \ref{sec:introduction}. A particular approach is to increase the number of times the light in the guided field interacts with the atomic dipole, as Eq. (\ref{eq:coop}) describes. This can be achieved by an optical cavity that recycles the field, effectively increasing the atom-light interaction time. It is possible to fabricate in-fiber cavities by modifying the optical fiber geometry to create Bragg mirrors that reflect the guided mode. Modified-ONFs are promising platforms for interfacing atoms and guided light in the strong coupling regime.

\subsubsection*{Photonic crystal cavities}

\begin{figure}
\centering
\includegraphics[width=0.75\textwidth]{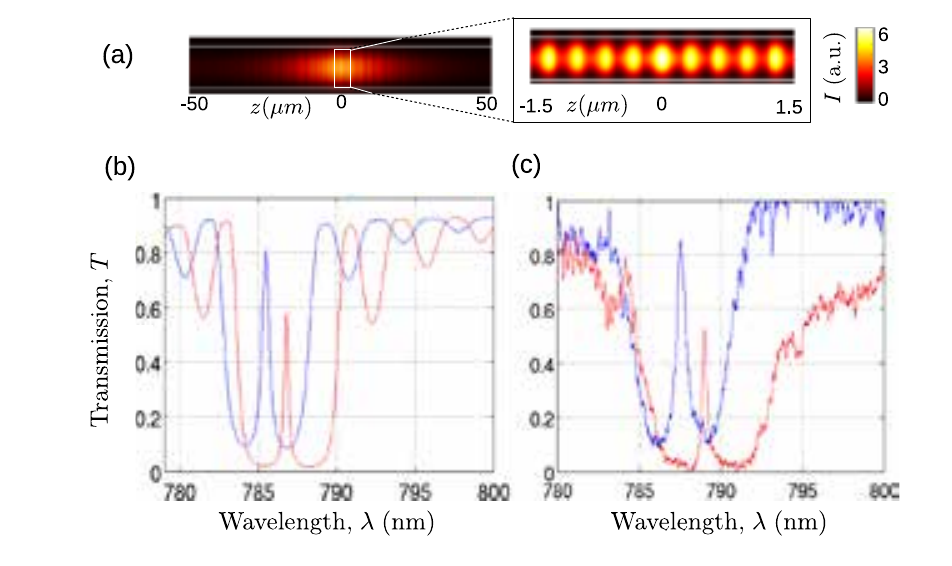}
\caption{\label{fig:Fig19}
(a) Simulated electric field intensity
inside the CPCC for the y-mode. The inset shows an expanded
view of the central region of the cavity. White horizontal lines
mark the position of the edges of the nanofiber. (b),(c) Simulated
and measured transmission spectra, respectively, for the x (blue
line) and y-modes (red line). (Reprinted Fig. 2 with permission from Yalla, R., Sadgrove, M., Nayak, K.P., Hakuta, K., 2014. Cavity quantum electrodynamics on a nanofiber using a composite photonic crystal cavity. Phys. Rev. Lett. 113 (14),143601, with Copyright (2014) by the American Physical Society.
of ~\cite{Yalla2014}.)}
\end{figure}

The evanescent field of a guided mode can be used not only for interacting light with single atoms but also with microscopic structures. A particular example is a photonic crystal in a Bragg mirror configuration. \cite{Sadgrove2013} and \cite{Yalla2014} implement an in-nanofiber Fabry-Perot-type cavity, coupling the evanescent field of the ONF guided mode to a photonic crystal structure by bringing them in contact. Fig. \ref{fig:Fig19} shows an example of the narrow-bandwidth transmission achieved with this technique by \cite{Yalla2014}. They observe quality factors better than 2000, offering a non-invasive method to create in-fiber cavities to enhance atom-light coupling.

\subsubsection*{Bragg resonators}

\begin{figure}
\centering
\includegraphics[width=0.45\textwidth]{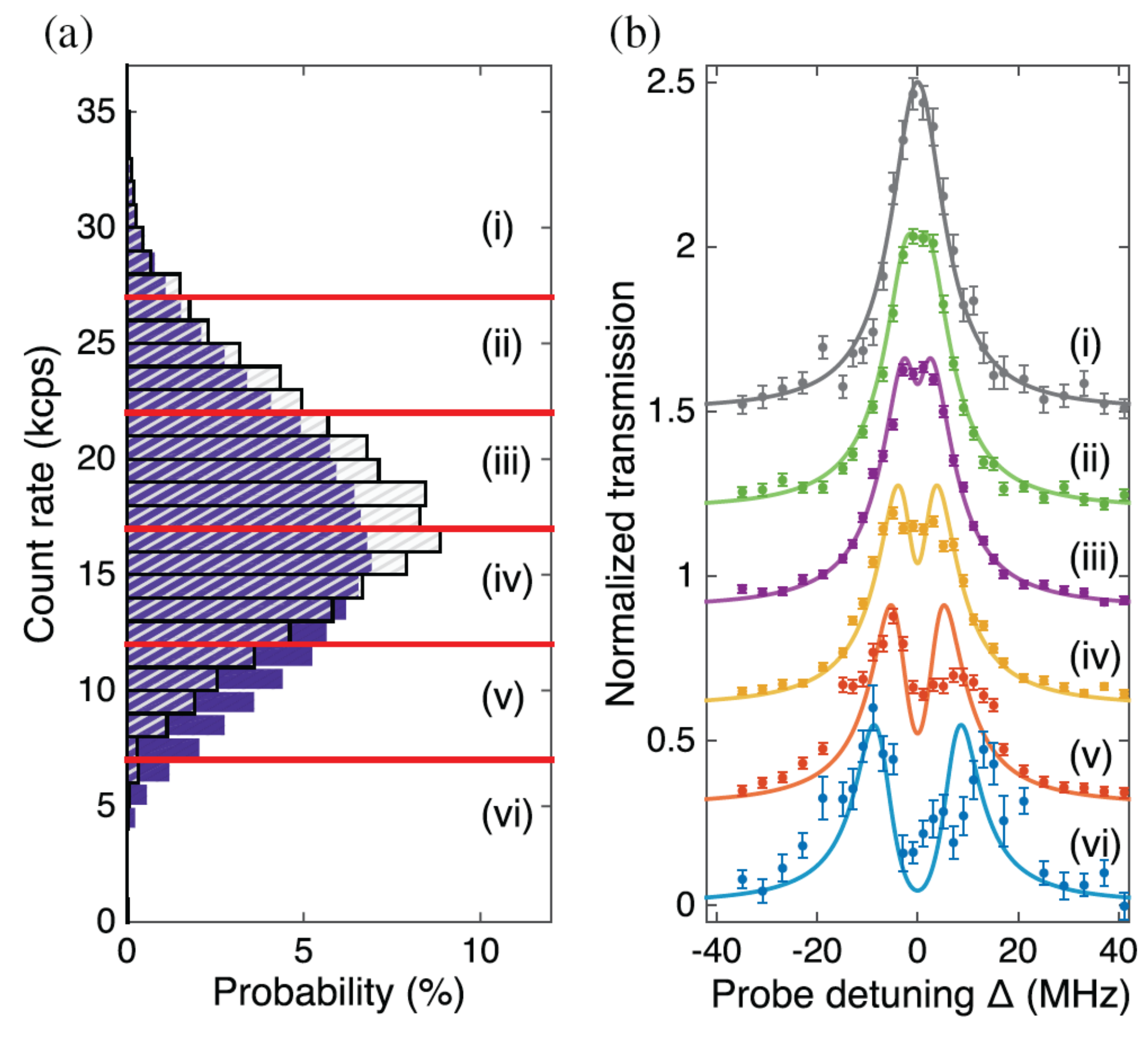}
\caption{\label{fig:Kato}
(a) Histogram of the transmission
intensity of the detection probe with (blue) and without (gray)
the optical molasses. The dark count of the detector is 1.4$\times10^3$
counts per second (cps), and the measurement background noise
including the dark count is about 10 kcps. (b) Transmission
spectra as functions of the probe detuning $\Delta$. Data sets and fits are
normalized to the empty-cavity transmission and are vertically
offset for clarity. The observed asymmetry in the spectra is
presumably due to the effect of the probe pulse on the center-of-mass
motion of the atom. Error bars are the standard error of the
mean. The atom-cavity coupling rates g for the fits are $2 \pi \times$(1.3; 1.9; 2.9; 4.3; 7.8) MHz for (ii) to (vi), respectively. (Reprinted Fig. 2 with permission from Kato, S., Aoki, T., 2015. Strong coupling between a trapped single atom and an all-ber cavity. Phys.
Rev. Lett. 115, 093603, with
Copyright (2015)
by the American Physical Society.
of ~\cite{Kato2015}).}
\end{figure}

The refractive index of an optical fiber can be modified using nanosecond pulses of UV-light in a highly germanium doped photosensitive fiber. A spatially-dependent refractive index can be imprinted in the fiber by modulating the intensity pattern of the writing laser. \cite{Lindner2009} use a Talbot interferometer to create a periodic modulation of the refractive index in a Bragg mirror configuration. Two Bragg reflectors produce an in-fiber Fabry-Perot-type optical cavity. \cite{Wuttke2012, Kato2015} use an optical fiber with printed Bragg reflectors and taper it with the flame-brushing technique to integrate an ONF inside an in-fiber optical cavity. Figure \ref{fig:Kato} shows the observation of vacuum Rabi splitting in this platform using a single trapped Cs atom in a state insensitive trap. This transmission measurement is evidence of atom-light interaction in the strong coupling regime where the coherent coupling is larger than the incoherent losses through spontaneous emission or cavity loss.

Bragg reflectors can also be implemented at the ONF waist by making a periodic array of holes in the fused silica nanofiber, like photonic crystal cavities. This has been achieved by photon ablation using femtosecond pulses of UV-light \cite{Nayak2012, Nayak2014}. Using this configuration, \cite{Nayak2014} report the fabrication of in-fiber cavities with a finesse up to 500. Another approach has been to use a focused ion beam milling technique for drilling periodic nano-grooves in the ONF \cite{Nayak2011}. This is the same method used in \cite{Daly2014,Daly2016} to create a slot at the ONF waist to enhance the interaction of the guided mode with particles inside the carved space.

\newpage
     %24 Dec 2016 V6 Luis
% Luis in charge 2017 01 17
% 2017-01-09 Pablo made the changes from the comments
% Luis continues to make changes 2017 01 09
%Luis, Fredrik,  and Pablo corrections
\section{Chirality in nanophotonic systems}
\label{sec:chirality}
The existence of longitudinal components of the electromagnetic field is well-established in ONFs (see Sec.~\ref{sec:modes} for a discussion of the polarization). Such polarizations also exist in the microwave regime and have been thoroughly used to make directional couplers and other important technological devices \cite{BadenFuller1986}. There, the waveguides are accompanied by materials with the appropriate electro-magnetic properties, that together with the geometry of the guides ensure the directionality. This enables microwave isolators that protect delicate hardware pieces, but also direct information and even quantum information to sensitive detectors.

\begin{figure}
\centering
\includegraphics[width=0.75\textwidth]{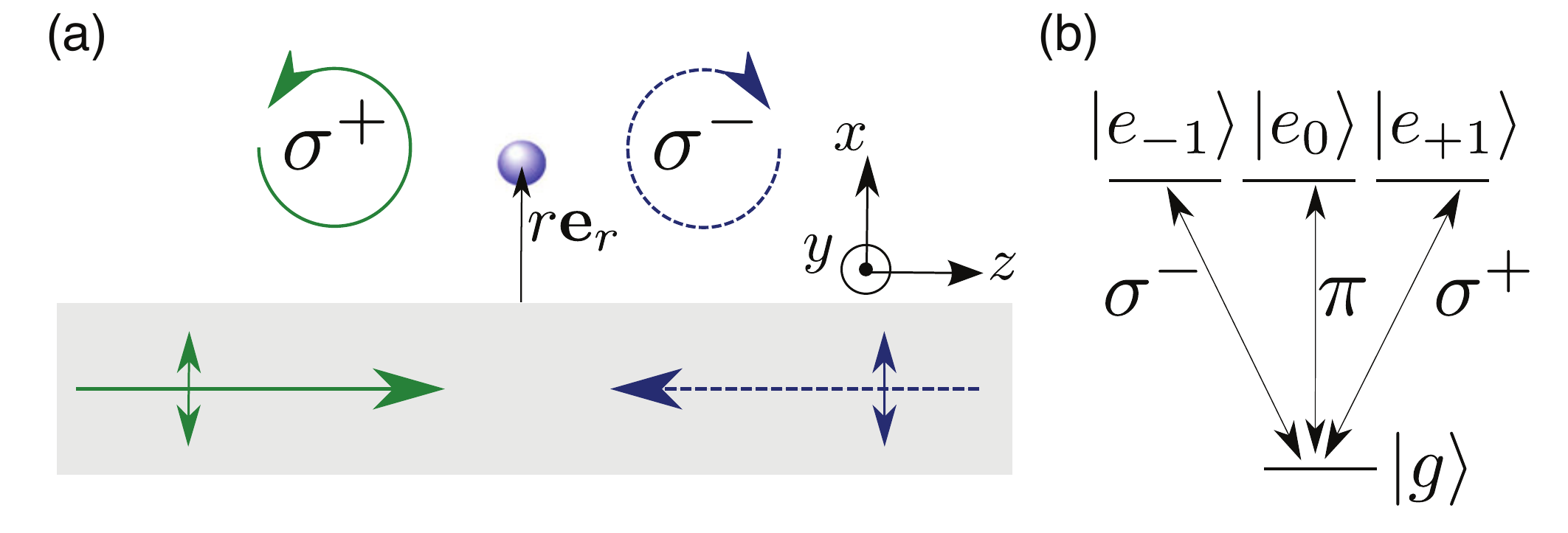}
\caption{\label{fig:chiral} Chiral photons in evanescent fields coupled to spin-polarized
atoms. (a) Polarization properties of the evanescent
light field that surrounds an optical nanofiber (gray). A light field
that propagates in the $(+z)$ direction and whose main polarization
axis (double arrow) is along the $x$ axis is almost fully $\sigma^{+}$ polarized
(green solid arrows) in the ($y=0$) plane. If it propagates in the
$(-z)$direction, it is almost fully $\sigma^{-}$  polarized (blue dashed
arrows). The quantization axis is chosen along y, i.e., orthogonal
to the propagation direction. An atom (light blue sphere) placed at
a distance r to the nanofiber surface couples to the evanescent
field. (b) Relevant energy levels of the atom. The ground state $\ket{g}$ is coupled to the excited states $\ket{e_{-1}}, \ket{e_{0}}$, and $\ket{e_{1}}$ via $\sigma^{-}$, $\pi$,
and $\sigma^{+}$ transitions, respectively. 
(Reprinted Fig. 1 with permission from Sayrin, C., Junge, C., Mitsch, R., Albrecht, B., O'Shea, D., Schneeweiss, P., Volz, J., Rauschenbeutel, A.,
Dec 2015b. Nanophotonic optical isolator controlled by the internal state of cold atoms. Phys. Rev. X 5,
041036.Open access journal~\cite{Sayrin2015a}).}
\end{figure}

Atoms are excellent materials for manipulating the directionality of electromagnetic waves through the chiral coupling of their momentum to the spin angular momentum of the light. ONFs with trapped atoms are becoming an important platform for chirality. Now it is possible to talk about chiral quantum optics, as the title of the recent review by \cite{Lodahl2016} is called. (See Fig.~\ref{fig:chiral} for a ONF setup). We point the interested reader to that reference as 
it covers this area in more detail and makes the important
connection with many-body physics, as it may become possible to realize interesting Hamiltonians as suggested by \cite{Ramos2014} and many others. The mode structure of the ONF also opens the platform to the rich spin-orbit interactions of light \cite{Bliokh2015}.  The intrinsic asymmetry of the problem is a new important tool in the harnessing of the atom-light interaction for further studies of quantum optics and quantum information.

\subsection{Optical control of directionality in photonic systems}

Realization of directionality through use of polarization selection rules, internal structure of the atoms, waveguide structures that spin-orbit couple the light, and chiral effects in nanophotonic systems has happened in different platforms: quantum dots in a cross waveguide, atoms in whispering-gallery-mode microresonator (WGM) and gold nanoparticles and atoms on an ONF.  They are an important starting point for what promises to be an active area of research. 

Crucial to the quantum dot work are first the highly polarized emission of its excitons using optical pumping, and second the emission properties of the quantum dots in a crossed-waveguide device with broken inversion symmetry. Together they enable optical control of the emission direction in \cite{Luxmoore2013}.

\cite{Junge2013} realize in a WGM microresonator  strong coupling between single atoms and nontransversal  photons. They profit from the fundamental change that the longitudinal polarization brings to the interaction between light and matter in this resonator.  Contrary to a macroscopic Fabry Perot, here the two
propagation directions of the photons are correlated with
two nearly orthogonal polarization states. As a consequence,
counterpropagating photons are distinguishable by their
polarization and cannot interfere destructively. 

\cite{Shomroni2014} and \cite{Scheucher2016} use the directional coupling and its dependence on the internal atomic state to create a single atom switch and router. Helped by an optical circulator (a WGM) evanescently coupled to an ONF and to a single atom, they are able to change the transmission and reflection of a single photon propagating through the nanofiber by preparing the state of the atom with a single triggering photon. Applications like these boost the role of ONF in photonic-based quantum technologies.

\subsection{Chirality in ONF systems}

\cite{LeKien2004OC} states that under certain circumstances if the
transverse component of the field is linearly polarized then the phase of the longitudinal component  differs from that of the transverse components by $\pm \pi/2$ depending on the direction of propagation, as explained in Eq.~\ref{eq:chirality} and Fig.~\ref{fig:chiral}. Due to this phase difference, the total electric field {\bf{E}} rotates elliptically with time, in a plane parallel to the fiber axis $z$.
This observation invites extending the tools used in the atom-light interaction in ONFs to include the longitudinal component of the field into the chiral domain.
\cite{Mitsch2014} 
demonstrate the directional spontaneous emission of photons by trapped Cs
atoms into a ONF controlling their asymmetric propagation direction with the excited state
of the atomic emitters. 

\cite{Petersen2014} control the flow of light with nanophotonic waveguides with the help of the spin-orbit coupling of light, and gold nanoparticles attached to an ONF. The nanoparticle breaks the mirror
symmetry of the scattering of light realizing a chiral waveguide coupler in which the handedness
of the incident light determines the propagation direction in the ONF. 
They achieve directionality of the scattering process up to 94\% of the in-coupled
light into a given direction. 
 
\cite{Sayrin2015a} realize a nanophotonic optical isolator with high optical isolation at the single-photon level. The direction of the resulting optical isolator is controlled by the spin
state of cold Cs atoms. The basic photonic and atomic structure is in Fig.~\ref{fig:chiral}. They perform measurements  with an ensemble of trapped
cold atoms where each atom is weakly coupled (a few percent for the reasons explained in Secs.~\ref{sec:around} and~\ref{sec:trapping}) to the ONF. To significantly enhance the coupling and be able to work with a single atom, they strongly couple the atom to a WGM
bottle microresonator similar to what they use for \cite{Junge2013}. They  observe simultaneously high isolation and high forward
transmission. 

We expect in the near future increasing use for technology such as isolators, but also the system is set to address questions in cascaded quantum open systems \cite{Carmichael1993, Gardiner1993} that have been open for more than two decades. Early work as that attempted by some of us  \cite{Gripp1995} could not reach the important couplings nor the directionality that the recent experiments have advanced, inviting exploration of the cascaded open systems and other questions related to chiral reservoirs for atoms around an ONF \cite{Ramos2014,Pichler2015,Guimond2016,Eldredge2016}.

\newpage
     
\section*{Conclusions}
\label{sec:conclusions}
ONF platforms with atoms are becoming an important player in the development of quantum optics and quantum information at the beginning of the XXI century. The structure has a simplicity and elegance that attracts many users. The large community that has grown using them has been able to control many of its technical challenges. Atoms are now routinely trapped around the nanofibers and provide robust platforms for quantum optics experiments. This would not have been possible without the many studies theoretical and experimental with atoms, both hot and cold, around the nanofibers.
These arrays of atoms around the ONF not only enable the exploration of the many-body physics of condensed matter system but given the peculiarities and interactions that are available, may allow for the study of Hamiltonians that do not exist in current materials.
The atom-atom interaction could in principle be infinite range, but it remains to be seen if the coherence times of the different ONF modes permit long range interactions. This may be possible as atoms around an ONF could have a phase imprinted without concern of any boundary conditions. The guided mode of the nanofiber manages to propagate without diffraction, maintaining a tight cross sectional area that can interact with many atoms along the way. Most work to date is with the fundamental mode, but there are many other modes available that also do not diffract and could be used. There are still some challenges and difficulties to be solved in the future, but the versatility of the three field polarizations available and the possibility to use the spin-orbit coupling of the light together with single atoms is an open avenue that will lead to many surprises and discoveries. We expect that parallel to the quantum optics and quantum information advances there will continue to be many technological applications that could have the potential to affect future areas of telecommunications. ONFs are a beautiful, versatile, and fun platform for future experiments in quantum optics.

\newpage
       % 2017 01 08 Luis
\section*{Acknowledgments}
This work has been supported by
%Army Research Office (ARO) (Atomtronics MURI (528418)); Defense Advanced Research Projects Agency (DARPA) (HR0011411122); 
National Science Foundation of the United States (NSF) (PHY-1307416);  NSF Physics Frontier Center at the Joint quantum Institute (PHY-1430094). S. R. acknowledges support from Fulbright
Foundation.
%Office of Naval Research (ONR).

%\end{document}

     \newpage
     \pagenumbering{arabic}
\bibliography{AAMOPreview.bib}

\end{document}